\documentclass[aps,amsmath,amssymb,showpacs,showkeys]{revtex4-2}
\usepackage[dvips]{graphicx,color}
\usepackage{times}
\usepackage{braket}
\usepackage{xcolor, soul}
\usepackage[
  colorlinks=true,
  urlcolor=blue,
  linkcolor=red,
  citecolor=blue
]{hyperref}
\usepackage{graphicx}
\usepackage{amsmath}
\usepackage{hyperref}
\hypersetup{colorlinks,citecolor=blue}
\hypersetup{colorlinks=true, linkcolor=red, filecolor=magenta, urlcolor=blue}
\usepackage{amsmath}
\usepackage{xcolor}
\usepackage{orcidlink}
\begin{document}
\title{Phase space analysis of Bianchi III Universe with $\boldsymbol{f(R,T)}$ gravity theory}

\author{Pranjal Sarmah\orcidlink{0000-0002-0008-7228}}
\email[E-mail:]{p.sarmah97@gmail.com}

\author{Umananda Dev Goswami\orcidlink{0000-0003-0012-7549}}
\email[E-mail:]{umananda@dibru.ac.in}

\affiliation{Department of Physics, Dibrugarh University, Dibrugarh 786004, 
Assam, India}

\begin{abstract}
The Bianchi type III (BIII) metric is a useful geometry to study cosmic 
anisotropies. It includes an extra exponential term multiplied by a directional 
scale factor and recasts the cosmological model as a dynamical system to 
provide various significant information regarding the evolution, stability of 
the system, etc. In this study, we have constructed a dynamical 
system for the BIII metric using $f(R,T)$ gravity theory and performed fixed 
point analysis in three different $f(R,T)$ models. Here, we have found that 
the first two  models i.e.~$f(R,T) = \alpha R + \beta f(T)$  and 
$f(R,T) = R + 2 f(T)$ are agreed  with standard $\Lambda$CDM cosmology but 
the third one i.e.~$f(R,T) = (\zeta+ \eta\, \tau\, T)R$ has the issue 
of unbounded energy density. Thus, we can remark that some $f(R,T)$ models 
may not be suitable for studying the evolution of the Universe with an 
anisotropic background, like using BIII metric, etc. However, all three models 
agree with the heteroclinic path of radiation-dominated, 
matter-dominated, and dark energy-dominated phases of the Universe as 
predicted by standard cosmology.
 
\end{abstract}


\maketitle

\section{Introduction}\label{1}
Homogeneity and isotropy are the two basic cosmological principles that form 
the basis of standard cosmology, often known as the $\Lambda$CDM cosmology. 
This cosmological model uses the Friedmann-Lema\^itre-Robertson-Walker (FLRW) 
metric and the  energy-momentum tensor in the standard form of a perfect 
fluid, to address most questions about the Universe \cite{Pebbles_1994}.
Researchers are encouraged to expand this traditional formalism to identify 
alternative theories or modifications to general relativity (GR) for a number 
of reasons, including the Universe's current rapid expansion 
\cite{Riess, Perlmutter_1999, Ma_2011}, zero observational evidence of dark 
matter (DM) \cite{Trimble_87} and dark energy (DE) \cite{Frieman_2008}. 
Therefore, a good number of modified formalisms of GR have been developed to 
disprove the notion of the Universe's exotic matter and exotic energy contents 
used to describe, respectively, its missing mass and current rapid expansion. 
These formalisms are generally known as the modified theories of gravity 
(MTGs).
  
One of the {simplest MTGs} is the $f(R)$ gravity theory 
\cite{Felice_2010,Faraoni_2010, Harko_2018} in which  Ricci scalar has been 
substituted by a function of the Ricci scalar $f(R)$ in the to modify the
Einstein-Hilbert action. This theory shows its versatility in cosmology, 
physics related to black holes \cite{Maroto_2009,Nashed_2022, Hazarika_2024,
Chaturvedi_2023, Karmakar_2024}, physics of the cosmic rays \cite{SPS_20241, 
SPS_20242}, and other theoretical research fields \cite{Gogoi_2020, Calza_2018, Ying_2014}. The Refs.~\cite{Gogoi_2021, Li_2007, Duniya_2023, 
Bajardi_2022,Goswami_2014} contain a comprehensive collection of cosmological 
studies for both conventional isotropic as well as anisotropic metrics using 
the $f(R)$ gravity theory. In $f(R,T)$ gravity theory, the gravitational 
Lagrangian density is composed of any arbitrary linear or nonlinear function 
of the Ricci scalar with the trace $T$ of the energy-momentum tensor 
$T_{\mu\nu}$\cite{Harko_2018, Harko_2011}. In this theory, the variation of 
the energy-momentum tensor with the metric is the source term, and it is 
represented by the term $\mathcal{L_M}$, which is known as the matter 
Lagrangian density in the general mathematical expression of the action. 
The field equations generated by each $\mathcal{L_M}$ option are distinct 
\cite{Kavya_2022, Jaybhaye_2024, Jaybhaye_2022}. Numerous studies have been 
conducted using the $f(R,T)$ gravity theory, some of them are included in the 
Refs.~\cite{Harko_2011, Jeakel_2024, Baffou_2021, Tretyakov_2018, Malik_2024, 
Rudra_2021}. In addition to these MTGs, some other theories, such as 
teleparallel gravity theories, like $f(T)$ theory of gravity 
\cite{Cai_2016, Sabiee_2022, Dimakis_2024, Fayaz_2014,Briffa_2023}, 
$f(Q)$ theory of gravity \cite{Sarmah_2023, Sarmah_2024, Khyllep_2023, espo, 
Koussour_2023, Shabani_2024, Shahoo_2024}, along with some 
standard model extension (SME), like the bumblebee gravity theory 
\cite{Capelo_2015, Maluf_2021, Sarmah_2024b}, which are usually termed 
together as the alternative gravity theories (ATGs) have gained popularity 
among cosmological scholars in recent years. Comparison of these MTGs and ATGs 
to data has yielded encouraging results \cite{Kavya_2022, Rudra_2021, 
Briffa_2023, Sarmah_2024b, Shahoo_2024}.
         
There are several shortcomings in the common cosmological principles of 
isotropy and homogeneity, in addition to the absence of direct observational 
proof of the existence of exotic matter and energy contents as mentioned 
earlier. Despite these presumptions, researchers have been able to explain the 
majority of cosmological aspects using the $\Lambda$CDM model, primary among 
them are the Hubble tension \cite{Planck_2018, Nedelco_2021}, the coincidence 
problem \cite{Velten_2014} and the $\sigma_8$ tension \cite{Planck_2018}. 
Nonetheless, many reliable observational data sources, such as WMAP 
\cite{wmap,wmap1,wmap2}, SDSS (BAO) \cite{SDSS_2005,Bessett_2009,Tully_2023}, 
and Planck \cite{Planck_2015,Planck_2018}, have revealed some departures 
from the conventional cosmological principles and raise the possibility 
that the Universe contains some anisotropies. Additionally, studies indicate 
that the Universe's shape is large-scale planar symmetric. In the CMB angular 
power spectrum, the eccentricity of order $10^{-2}$ can match the quadrupole 
amplitude with observational data without altering the higher-order multipole 
of temperature anisotropy \cite{Tedesco_2006}. The existence of 
asymmetry axes in the Universe is confirmed by polarization studies of 
electromagnetic radiation traveling great distances \cite{akarsu_2010}. Thus, 
not all cosmic features can be adequately explained by the principles of 
homogeneity and isotropy alone.
 
A metric with an anisotropic background and homogeneous in nature is 
required to understand the anisotropic characteristics of the current 
Universe. In order to extract anisotropy information for cosmological 
investigations, researchers often choose Bianchi metrics known as types 
I (BI), III (BIII), V (BV) and IX (BIX), which are among the eleven types of 
anisotropic metrics that Luigi Bianchi suggested \cite{Riyan_1975}. The 
majority of the cosmological studies in these metrics are mostly confined to 
type I, and relatively little research has been done in this area. Some 
works on anisotropic cosmology based on BI metric can be found 
in Refs.~\cite{Sarmah_2022,Cea_2022, Perivolaropoulos_2014, Berera_2004, 
Campanelli_2006, Campanelli_2007, Paul_2008, Barrow_1997, akarsu_2019}. 
However, as mentioned already the study based on the BIII metric 
is very much limited. Apart from the exponential term contained in the 
metric, BIII has some other interesting features like the homogenous 
background \cite{Sahoo_2016, Riyan_1975,Katore_2016, Mollah_2018}, isotropic with time 
\cite{Katore_2016, Mollah_2018} and reducibility to the BI metric 
\cite{Sarmah_24c}. Again, the metric possesses hyper-spherically curved 
geometry \cite{Singh_2023}. All these characteristics have made this model 
interesting for cosmological studies. A few works in the BIII metric are 
mentioned here as follows. A.~Moussiaux et al.~in 1981 had found an exact 
solution to the Einstein field equations (EFE) in vacuum with a cosmological 
constant in BIII geometry \cite{Mouss_1981}. D.~Lorenz had proposed a model 
that incorporates both dust matter and the cosmological constant in the BIII 
geometry in 1982 \cite{Lorenz_1982}. Ref.~\cite{chak_2001} has made another 
study that suggested a viscous model of cosmology with the variable 
cosmological constant ($\Lambda$) and the gravitational constant ($G$). 
J.~P.~Singh et al.~had examined a work with the variable gravitational 
constant $G$ and cosmological constant $\Lambda$ in perfect fluid, assuming a 
conservation rule for the energy-momentum tensor ($T_{\mu\nu}$) in the BIII 
geometry in 2007 \cite{Singh_2007}. In Ref.~\cite{tiwari_2009}, a BIII model 
with the conventional perfect fluid and the time-dependent cosmological 
constant ($\Lambda$) along with constant deceleration parameter were studied. 
P.~S.~Letelier in 1980 had studied several two-fluid cosmological 
models that have similar symmetries like the BIII models. Here, in these 
models, an axially symmetric anisotropic pressure is produced by the 
independent four velocity associated with  the two non-interacting perfect 
fluids \cite{Lete_1980}.

In this present work, we try to study the Universe with BIII geometry as a 
dynamical system to understand its evolution and the possibility of various 
phases of it. Constructing Universe as a Dynamical system from cosmological 
equations and performing phase space analysis is a powerful and elegant way 
to study the dynamics of the Universe \cite{Agostino_2018}. This technique 
has been used to study the stability of systems employing a different models 
of cosmology including the both canonical \cite{Heard_2002,Fang_2014} and 
non-canonical models of the scalar-fields \cite{Fang_2016, Copeland_2005, 
Copeland_2010}, the theories of scalar-tensor \cite{Agarwal_2008,Huang_2015}, 
the $f(R)$  theory of gravity \cite{Amendola_2007,Carloni_2015, Alho_2016}, 
and others \cite{Odintsov_2018, Li_2018, Odintsov_2017,Hrycyna_2013}. 
Ref.~\cite{Bahmonde_2018} provides a contemporary overview of the 
phase space analysis techniques through constructing dynamical 
system from cosmological equations. A study based on this method for isotropic 
Universe in $f(R,T)$ theory has been found in Ref.~\cite{Mirza_2016}. 
Similar studies for the cosmic dynamical system are also carried out in $f(Q)$ 
gravity theory and they have been found in Refs. \cite{De_2023, Lu_2019}. 
However, there aren't many comparable studies using the Bianchi models in the 
$f(R,T)$ gravity theory. Thus we intend to investigate it using the BIII 
metric to learn more about the anisotropic Universe in $f(R,T)$ gravity 
theory. Here we are considering the BIII geometry in $f(R,T)$ theory of 
gravity to understand the Universe as a dynamical system. 
In this work, we are using the constrained values of various cosmological 
parameters from Ref.~\cite{Sarmah_24c} which are constrained through various 
observational data includes  Hubble data, data of SNe Ia (Pantheon plus data), 
BAO data, etc., through the powerful Bayesian inference technique, and plot 
the phase portraits of dynamical variables to understand the evolutionary 
track of the Universe for different $f(R,T)$ models.
   
This article has been structured as follows. The introduction 
part in Section \ref{1} briefly explains the importance of the BIII Universe 
and MTGs, specifically the $f(R,T)$ theory of gravity from various 
literature. In Section \ref{2}, we have discussed the general form of 
field equations in $f(R,T)$ gravity theory. In Section \ref{3}, we have 
developed the required field equations and continuity equation in the BIII 
metric for two different $f(R,T)$ models. In Section \ref{4}, we have performed 
the dynamical system analyses test for {three} different $f(R,T)$ 
gravity models and tested their physical compatibility. Finally, the article 
has been summarized with conclusions in Section \ref{6}.

\section{$\boldsymbol{f(R,T)}$ gravity theory and field equations}\label{2}

As mentioned in the previous section, the $f(R,T)$ theory of gravity 
introduces the functional term containing  $T$, which is the trace of 
$T_{\mu\nu}$ (i.e. the energy-momentum tensor) and the Ricci Scalar ($R$). 
This dependence of $T$ may be introduced by considering the conformal anomaly 
due to exotic imperfect fluid or quantum effect etc.~\cite{Harko_2011}. By 
choosing a proper choice of the function of $T$, one can reconstruct 
arbitrary FLRW cosmology or 
standard cosmology, without considering conventional classical GR theory 
\cite{Harko_2011}. Thus, for the $f(R,T)$ gravity theory, the modified 
Einstein-Hilbert (E-H) action may  be expressed as \cite{Harko_2011}
\begin{equation}\label{action}
{S} = \frac{1}{2\kappa} \int {f(R,T)} \sqrt{-{g}} ~ {d^{4}} x + \int {\mathcal{L_M}} \sqrt{-{g}} ~ {d^{4}x},
\end{equation}
where $\kappa = 8\pi G$. The field equations for the theory can be obtained by 
varying the modified action \eqref{action} with the metric tensor 
$g_{\mu\nu}$ and can be written as
\begin{align}\label{FE}
f_{R}(R,T)R_{\mu \nu} - \frac{1}{2}f(R,T)\,g_{\mu \nu} + ( g_{\mu \nu} \square- \nabla_{\mu}\nabla_{\nu})f_{R}(R,T)=\kappa\, T_{\mu\nu}-f_{T}(R,T)\,T_{\mu\nu} - f_{T}(R,T)\,\Theta_{\mu \nu},
\end{align}
where $f_{R}(R,T)$ and $f_{T}(R,T)$ are the derivatives of $f(R,T)$
with respect to Ricci scalar $R$ and trace of the energy-momentum tensor $T$ 
respectively, and the energy-momentum tensor (EMT) $T_{\mu\nu}$ can be 
written as
\begin{equation}\label{BE}
T_{\mu \nu} = -\,\frac{2}{\sqrt{-g}}\frac{\delta \left( \sqrt{-g}\,\mathcal{L_M}\right)}{\delta g^{\mu \nu}}.
\end{equation}
Further, the term $\Theta_{\mu \nu}$ of equation \eqref{FE} can be expressed as \cite{Harko_2011}
\begin{equation}
\Theta_{\mu \nu} = -\,2\,T_{\mu\nu} + g_{\mu\nu} \mathcal{L_M}-2\,g^{\alpha\beta} \frac{\partial^{2}\mathcal{L_M}}{\partial g^{\mu\nu} \partial g^{\alpha \beta}}.
\end{equation}
For the conventional diagonal energy-momentum tensor ($T_{\mu}^{\nu}$), 
the considered matter Lagrangian density has the form $\mathcal{L_M}=-\,p$ 
\cite{Harko_2011}. Thus, the $\Theta_{\mu\nu}$ reduces to 
\begin{equation}\label{theta}
\Theta_{\mu\nu} = -\,2\,T_{\mu\nu}-p\,g_{\mu\nu}.
\end{equation}

With the help of the field equations \eqref{FE}, we have written the 
field equations for the BIII metric using the conventional perfect fluid in 
the next sections for some selected $f(R,T)$ models.

\section{Bianchi III Cosmology in $\boldsymbol{f(R,T)}$ gravity Theory}\label{3}
The general form of BIII metric is
\begin{equation}\label{metric}
ds^2 = -\,dt^2+a_1^2(t)\, dx^2+ a_2^2(t)\,e^{-2mx} dy^2+ a_3^2(t)\, dz^2.
\end{equation} 
Here, $a_1$, $a_2$ and $a_3$ are the  scale factors along $x$, $y$ and $z$ 
respective directions and the term $m$ is a constant. Thus, we have three  
Hubble parameters $H_1 =\dot{a_1}/a_1$, $H_2 = \dot{a_2}/a_2$  and 
$H_3 = \dot{a_3}/a_3$ for three different directions mainly along $x$, $y$ 
and $z$  respectively. Further, for the average scale factor 
$a = (a_1 a_2 a_3)^{\frac{1}{3}}$ for the  considered metric we can write  
the average Hubble parameter as

\begin{equation}\label{Hub}
H = \frac{1}{3}(H_1+H_2+H_3).
\end{equation}

In our work, we have taken the conventional perfect fluid from of the 
energy-momentum tensor which has the mathematical form:
$T_{\mu}^{\nu} = diag(-\,\rho,p,p,p)$, and therefore the components of the 
tensor $\Theta_{\mu\nu}$ from equation \eqref{theta} now can be written as
$\Theta_{00} = -2\rho + p$, $ \Theta_{11} = -3a_1^{2}\,p$, 
$ \Theta_{22} = -3a_2^{2}\,e^{-2mx}\,p$, 
$ \Theta_{33} = -3a_3^{2}\,p$ for $\mathcal{L_M}=-\,p$ \cite{Sarmah_24c}.
Now, we are ready to derive field equations for the $f(R,T)$ gravity 
models under consideration of the BIII metric. We have considered the 
following three forms of $f(R,T)$ models in our study as suggested in 
Ref.~\cite{Harko_2011}:
\begin{equation}
f(R, T) = 
\begin{cases} 
f_1(R) + f_2(T),\\[3pt]
R + 2f(T),\\
f_1 (R)+ f_2(R)f_3(T)
\end{cases}
\end{equation}
With these {three} forms of models, we proceed as follows.  
  
\subsection{$f(R,T) = f_1(R) + f_2(T)$}

This is a standard mathematical form of the $f(R,T)$ gravity model. In our 
study, we have considered the explicit form of this model as 
$f(R,T) = \alpha R + \beta f(T)$. Here $\alpha$, $\beta$ are two free model 
parameters. The generalized form of the field equations for the considered 
$f(R,T)$ model can be written as
\begin{equation}\label{FEn}
\alpha \Big(R_{\mu\nu} -\frac{1}{2}g_{\mu\nu}R\Big) = \big[\kappa + \beta f_{T}(T)\big]T_{\mu\nu} +\Big[\beta\, p f_{T}(T)+\frac{1}{2} \beta f(T)\Big]g_{\mu\nu} {=\kappa T^{eff}_{\mu\nu}},
\end{equation}
where $T^{eff}_{\mu\nu}$ be the effective energy-momentum tensor for the 
considered model and it can be written as
$T^{eff}_{\mu\nu} = T_{\mu\nu} + \tilde T^{\mu\nu}$ with
\begin{equation}\label{EMTt}
\kappa \tilde T^{\mu\nu} = \beta f_{T}(T)T_{\mu\nu} +\Big[\beta\, p f_{T}(T)+\frac{1}{2} \beta f(T)\Big]g_{\mu\nu}.
\end{equation} 
The field equations from equation \eqref{FEn} have been obtained for the BIII 
metric with $f(T) = \lambda T$, where $\lambda$ is a constant, and the perfect 
fluid energy-momentum tensor and they have the form \cite{Sarmah_24c}
\begin{align}\label{fet}
\frac{\dot{a_1}\dot{a_2}}{a_1 a_2}+\frac{\dot{a_2}\dot{a_3}}{a_2 a_3}+\frac{\dot{a_3}\dot{a_1}}{a_3 a_1}  - \left(\frac{m}{a_1}\right)^2 & = \frac{1}{\alpha}\left(\kappa + \frac{3}{2} \lambda \beta \right)\rho - \frac{5\lambda \beta}{2 \alpha}\, p,\\[5pt]
\label{fex}
\frac{\ddot{a_2}}{a_2}+\frac{\ddot{a_3}}{a_3}+ \frac{\dot{a_2}\dot{a_3}}{a_2 a_3} & = -\, \frac{1}{\alpha}\left(\kappa + \frac{7}{2} \lambda \beta \right)p + \frac{\lambda \beta}{2 \alpha}\, \rho,\\[5pt]
\label{fey}
 \frac{\ddot{a_3}}{a_3}+\frac{\ddot{a_1}}{a_1}+ \frac{\dot{a_3}\dot{a_1}}{a_3 a_1} & = -\, \frac{1}{\alpha}\left(\kappa + \frac{7}{2} \lambda \beta \right)p + \frac{\lambda \beta}{2 \alpha}\, \rho,\\[5pt]
\label{fez}
\frac{\ddot{a_1}}{a_1}+\frac{\ddot{a_2}}{a_2}+ \frac{\dot{a_1}\dot{a_1}}{a_1 a_2}-\left(\frac{m}{a_1}\right)^2 & = -\, \frac{1}{\alpha}\left(\kappa + \frac{7}{2} \lambda \beta \right)p + \frac{\lambda \beta}{2 \alpha}\, \rho,\\[5pt]
\label{fexy}
m\left(\frac{\dot{a_1}}{a_1}-\frac{\dot{a}_2}{a_2}\right) & = 0.
 \end{align} 
 
We observe from equation \eqref{fexy} that there are few possibilities 
regarding the parameter $m$ and the directional Hubble's parameters. For 
$m\neq 0$, $H_1 = H_2$  the field equations may be rewritten 
as follows \cite{Sarmah_24c}:
\begin{align}\label{fe1}
H_1^2 + 2 H_1 H_3 - \left(\frac{m}{a_1}\right)^2 & = \frac{1}{\alpha}\left(1 + \frac{3}{2} \lambda \beta \right)\rho - \frac{5\lambda \beta}{2 \alpha}\, p,\\[5pt]
\label{fe2}
H_1^2+ H_3^2 + H_1 H_3 + \left(\dot{H_1} + \dot{H_3} \right) & = -\,\frac{1}{\alpha}\left(1 + \frac{7}{2} \lambda \beta \right)p + \frac{\lambda \beta}{2 \alpha}\, \rho, \\[5pt]
\label{fe3}
3H_1^2+ 2\dot{H_1} -\left(\frac{m}{a_1}\right)^2 & = -\, \frac{1}{\alpha}\left(1 + \frac{7}{2} \lambda \beta \right)p + \frac{\lambda \beta}{2 \alpha}\, \rho.
\end{align}
Here we consider the geometrized unit system and thus consider 
$\boldsymbol{\kappa = 1}$ \cite{Harko_2018}.  On the other hand, for the condition $m=0$, 
equations \eqref{fet}, \eqref{fex}, \eqref{fey} and \eqref{fez} can be rewritten as \cite{Sarmah_24c}
\begin{align}
H_1 H_2+H_2 H_3 + H_3 H_1 & = \frac{1}{\alpha}\left(1 + \frac{3}{2} \lambda \beta \right)\rho - \frac{5\lambda \beta}{2 \alpha}\, p,\\[5pt]
H_2^2+ H_3^2 + H_2 H_3 + \left(\dot{H_2} + \dot{H_3} \right) & = -\,\frac{1}{\alpha}\left(1 + \frac{7}{2} \lambda \beta \right)p + \frac{\lambda \beta}{2 \alpha}\, \rho,\\[5pt]
H_1^2+ H_3^2 + H_1 H_3 + \left(\dot{H_1} + \dot{H_3} \right) & = -\,\frac{1}{\alpha}\left(1 + \frac{7}{2} \lambda \beta \right)p + \frac{\lambda \beta}{2 \alpha}\, \rho,\\[5pt]
H_1^2+ H_2^2 + H_1 H_2 + \left(\dot{H_1} + \dot{H_2} \right) & = -\,\frac{1}{\alpha}\left(1 + \frac{7}{2} \lambda \beta \right)p + \frac{\lambda \beta}{2 \alpha} \rho.
\end{align}
Further, for the condition of $m = 0$ and $H_1 = H_2$ , those field equations 
now can be rewritten as \cite{Sarmah_24c}
 \begin{align}
 H_1^2+ 2H_3 H_1 & = \frac{1}{\alpha}\left(1 + \frac{3}{2} \lambda \beta \right)\rho - \frac{5\lambda \beta}{2 \alpha}\, p,\\[5pt]
 H_1^2+ H_3^2 + H_1 H_3 + \left(\dot{H_1} + \dot{H_3} \right) & = -\,\frac{1}{\alpha}\left(1 + \frac{7}{2} \lambda \beta \right)p + \frac{\lambda \beta}{2 \alpha}\, \rho,\\[5pt]
3H_1^2 + 2\dot{H_1} & = -\,\frac{1}{\alpha}\left(1 + \frac{7}{2} \lambda \beta \right)p + \frac{\lambda \beta}{2 \alpha}\, \rho.
\end{align}
The metric equation \eqref{metric} shows that when $m = 0~\& ~H_1 = H_2$, the 
metric reduces to LRS-BI Universe from BIII Universe, and for $m=0$ it 
reduces to the standard BI Universe. We summarize all the above mentioned 
conditions on the basis of values of $m$ in Table \ref{tab1}.
\begin{center}
\begin{table}[!h]
\caption{Physical situations for different conditions on the parameter $m$.}
\vspace{5pt}

\begin{tabular}{|c|c|c|}
\hline
\rule[1ex]{0pt}{2.5ex} Conditions~~&$H_1$ \& $H_2$~~relation & Remarks \\[2pt] 
\hline
\rule[1ex]{0pt}{2.5ex}$m = 0$ & $H_1 \neq H_2$ & BI Universe \\ 
\rule[1ex]{0pt}{2.5ex} {$m = 0$} &$H_1 = H_2$ & LRS-BI Universe \\
\rule[1.25ex]{0pt}{2.5ex} $m \neq 0$& $H_1 = H_2$ & BIII Universe \\ 
\hline
\end{tabular}
\label{tab1}
\end{table} 
\end{center}
 
As our primary objective is to explore the BIII metric implications, hence we 
are only interested in the $m\neq 0~ \& ~H_1=H_2$ case in our study.  
Therefore, the shear scalar $\sigma^2$ can be written as
 
 \begin{equation}
 \sigma^2 = \frac{1}{3}\left(H_1-H_3\right)^2.
 \end{equation}
\subsubsection{Cosmological parameters for $m \neq 0~ \&~ H_1 = H_2$ case}
For $m \neq 0$ and $H_1 = H_2$, we have written the field equations 
as given equations \eqref{fe1}, \eqref{fe2} and \eqref{fe3}. Now further 
considering $H_3 = \gamma H_1$ through using the $\theta \propto \sigma$ 
condition in which $\theta^2$ and $\sigma^2$ as the expansion and shear scalar 
respectively. This condition of proportionality is a well-known 
condition in anisotropic cosmology, and it has been widely discussed in 
several literature and here we have mentioned a few in Refs.~\cite{Colins_1971, Colins_1980, Ban_1985, Ban_1988, Rib_1987}. Now, the compact form of the field equations have
\begin{align}
3H^2 & = \frac{(2+\alpha)^2}{3\alpha(1+2\gamma)} \left[(1+\frac{3}{2}\lambda \beta)\rho -\frac{5}{2}\lambda \beta p + \frac{\alpha m^2}{a^{\frac{6}{2+\gamma}}}\right]\label{Fen1},\\[8pt]
3H^2 + \frac{2}{3}\left(2+\gamma\right)\dot{H} & = -\,\frac{(2+\gamma)^2}{9\alpha}\left[(1+\frac{7}{2}\lambda \beta)p -\frac{1}{2}\lambda \beta \rho - \frac{\alpha m^2}{a^{\frac{6}{2+\gamma}}}\right].\label{Fen2}
 \end{align}
As the $\boldsymbol{T_{\mu\nu}}$ for the conventional perfect fluid is not
conserved in $f(R,T)$ theory of gravity  as shown in Refs.~\cite{O_2014, Carvalho_2020, 
Santos_2019, Moraes_2018}, thus the modified continuity equation for $f(R,T)$
theory of gravity can be written by using the condition of conservation
relation, 
$\nabla_{\mu}T^{\mu\nu}_{eff}= \nabla_{\mu}T^{\mu\nu}+
\nabla_{\mu}{\tilde T^{\mu\nu}} = 0$ \cite{Moraes_2018} as
\begin{equation}\label{cont}
\dot{{\rho}} = -\,\frac{3{H}\rho\left\{1-{{\lambda} {\beta}}\right\}(1+{\omega})}{\left\{1+\frac{{\lambda} {\beta}}{2}(5{\omega}-3)\right\}}.
\end{equation}
For the considered set of field equations, the components 
$\tilde{T}^{0}_{0} =\frac{\lambda \beta}{2}(-3\rho+5p)$, 
$\tilde{T}^{1}_{1} =\frac{\lambda \beta}{2}(-\rho+7p)$, 
$\tilde{T}^{2}_{2} =\frac{\lambda \beta}{2}(-\rho+7p)$ and
$\tilde{T}^{3}_{3} =\frac{\lambda \beta}{2}(-\rho+7p)$ can be obtained from 
the relation \eqref{EMTt}.

With the use of equations \eqref{Fen1} and \eqref{Fen2}, the effective 
equation of state can be written as
\begin{equation}\label{Om_eff}
\omega_{eff} = -\left( 1+\frac{2(2+\gamma)}{9}\frac{\dot{H}}{H^2}\right)=\frac{(2+\gamma)^2(1+2\gamma)}{3(2+\alpha)^2}\left[\frac{(1+\frac{7}{2}\lambda \beta)p -\frac{1}{2}\lambda \beta \rho - \alpha\, m^2 a^{-\frac{6}{2+\gamma}}}{(1+\frac{3}{2}\lambda \beta)\rho -\frac{5}{2}\lambda \beta p + \alpha\, m^2a^{-\frac{6}{2+\gamma}}}\right].
\end{equation}
Here, the effective equation state $\omega_{eff}$ means the ratio of the 
effective pressure to the effective density. Further, the  right hand side of 
equation \eqref{Fen1} can be considered to be the effective density term by 
comparing it with the  Friedmann equation for FLRW metric and similarly the 
right hand side of equation \eqref{Fen2} can be considered as the effective 
pressure term \cite{Sarmah_24c}. Accordingly $q$ i.e. the expression of 
deceleration parameter can be written as
 \begin{equation}\label{dec}
 q =-\left(1+\frac{\dot{H}}{H^2}\right) = \frac{(5-2\gamma)+9\,\omega_{eff}}{2(2+\gamma)}.
 \end{equation}
Again, using the relation $p = \omega \rho$ in which $\omega$ is the equation 
of state with $\omega = 0$,  $\omega = \frac{1}{3}$ and $\omega = -\,1$ for 
matter, radiation and  dark energy respectively. Now, the solution for $\rho$ 
for equation \eqref{cont} can be obtained as
\begin{equation}\label{rho}
\rho = \rho_0 (1+z)^{\frac{3(1+\omega)\{1-{\lambda \beta}\}}{\{1+\frac{\lambda \beta}{2}(5\omega-3)\}}}.
 \end{equation} 
 
Now, we are ready with this set of equations for further phase space  
analysis. The detailed mathematical formulations and methods for the same 
have been discussed in later sections. 

\subsection{$\boldsymbol{f(R,T) = R + 2 f(T)}$}
It is the reduced version of the previously considered 
$f(R,T) = \alpha R + \beta f(T)$ model. In this model the value of $\alpha$ 
and $\beta $ are $1$ and $2$ respectively. Further, we have considered  
$f(T) = \lambda T$. Thus, the reduced form of the field equations for this 
model can be written for $m\neq 0$, $H_1 = H_2$ and $H_3 = \gamma H_1$ as
\begin{eqnarray}
3H^2 = \frac{3}{(1+2\gamma)} \left[(1+3\lambda)\rho - 5\lambda p + \frac{ m^2}{a^{\frac{6}{2+\gamma}}}\right]\label{hub},\\[8pt]
3H^2 + \frac{2}{3}\left(2+\gamma\right)\dot{H}=-\frac{(2+\gamma)^2}{9}\left[(1+7\lambda)p - \lambda \rho - \frac{ m^2}{a^{\frac{6}{2+\gamma}}}\label{Fe2}\right].
\end{eqnarray}
Further, both the equations \eqref{cont} and \eqref{rho} are now can be 
written for this case as
\begin{align}\label{cont_2}
\dot{\rho} & = -\,\frac{3H\rho(1+\omega)\{1-{2\lambda}\}}{\{1+{\lambda}(5\omega-3)\}},\\[8pt]
\label{rho_2}
\rho & = \rho_0 (1+z)^{\frac{3(1+\omega)\{1-{2\lambda}\}}{\{1+{\lambda}(5\omega-3)\}}}.
\end{align} 
Consequently, the expressions of the $\omega_{eff}$ and $q$ for this model 
can be derived from equations \eqref{Om_eff} and \eqref{dec} respectively.
  
\section{Dynamical system analysis of the BIII metric in $\boldsymbol{f(R,T)}$ models}\label{4}
In this section, we analyse the BIII Universe as a dynamical system by using
the above mentioned  $f(R,T)$ gravity models. Here, we try to assess the system 
stability by identifying the critical points and their characteristics. 
\subsection{$\boldsymbol{f(R,T) = \alpha R + \beta f(T)}$ Model}
Using dynamical variables, the Friedmann like equation \eqref{Fen1} for this 
model can be expressed in the form:
\begin{equation}\label{Fried1}
 \frac{(2+\alpha)^2}{3\alpha(1+2\gamma)}\left[(1+\frac{3}{2}\,\lambda \beta)X + (1+\frac{2}{3}\, \lambda \beta)Y + (1+4\lambda\beta)Z + \alpha m^2 S  \right] = 1,
 \end{equation}
where $X = \rho_m/3H^2$, $Y = \rho_r/3H^2$, $Z = \rho_{\Lambda}/3H^2$ and 
$S = (3H^2 a^{6/(2+\gamma)})^{-1}$ are the dynamical variables with $\rho_m$,
$\rho_r$ and $\rho_{\Lambda}$ as the matter, radiation and dark 
energy densities respectively. $X$, $Y$ and $Z$ are the density 
parameters of matter ($\Omega_m$), radiation ($\Omega_r$) and dark energy 
($\Omega_{\Lambda}$) respectively. Further, we can rewrite 
the equation \eqref{Fen2} by considering the dynamical variables and the 
equation \eqref{Fried1} as
\begin{equation}\label{e2}
\frac{\dot{H}}{H^2} = -\,\frac{2+\gamma}{2\alpha}\left[(1+\lambda\beta)X + \frac{4}{3}(1+\lambda \beta)Y - 3\alpha\left\{\frac{1+2\gamma}{(2+\alpha)^2} - \frac{3}{(2+\gamma)^2}\right\} \right].
\end{equation}
Again, the derivatives $X' = \frac{dX}{dN}$ and $Y' = \frac{dY}{dN}$ with 
$N =\log a$  and for $S\ll1$ gives,
\begin{eqnarray}\label{xdynamic}
X' = \frac{2+\gamma}{\alpha}X \left[(1+\lambda\beta)X + \frac{4}{3}(1+\lambda \beta)Y - 3\alpha\left\{\frac{1+2\gamma}{(2+\alpha)^2} - \frac{3}{(2+\gamma)^2} + \frac{2(1-\lambda \beta)}{(2+\gamma)(2-3\lambda\beta)}\right\} \right],\\[8pt]
\label{ydynamic}
Y' = \frac{2+\gamma}{\alpha}Y \left[(1+\lambda\beta)X + \frac{4}{3}(1+\lambda \beta)Y - 3\alpha\left\{\frac{1+2\gamma}{(2+\alpha)^2} - \frac{3}{(2+\gamma)^2} + \frac{4(1-\lambda \beta)}{(2+\gamma)(3-2\lambda\beta)}\right\} \right].
\end{eqnarray}

\begin{figure}[!h]
\centerline{
  \includegraphics[scale = 0.42]{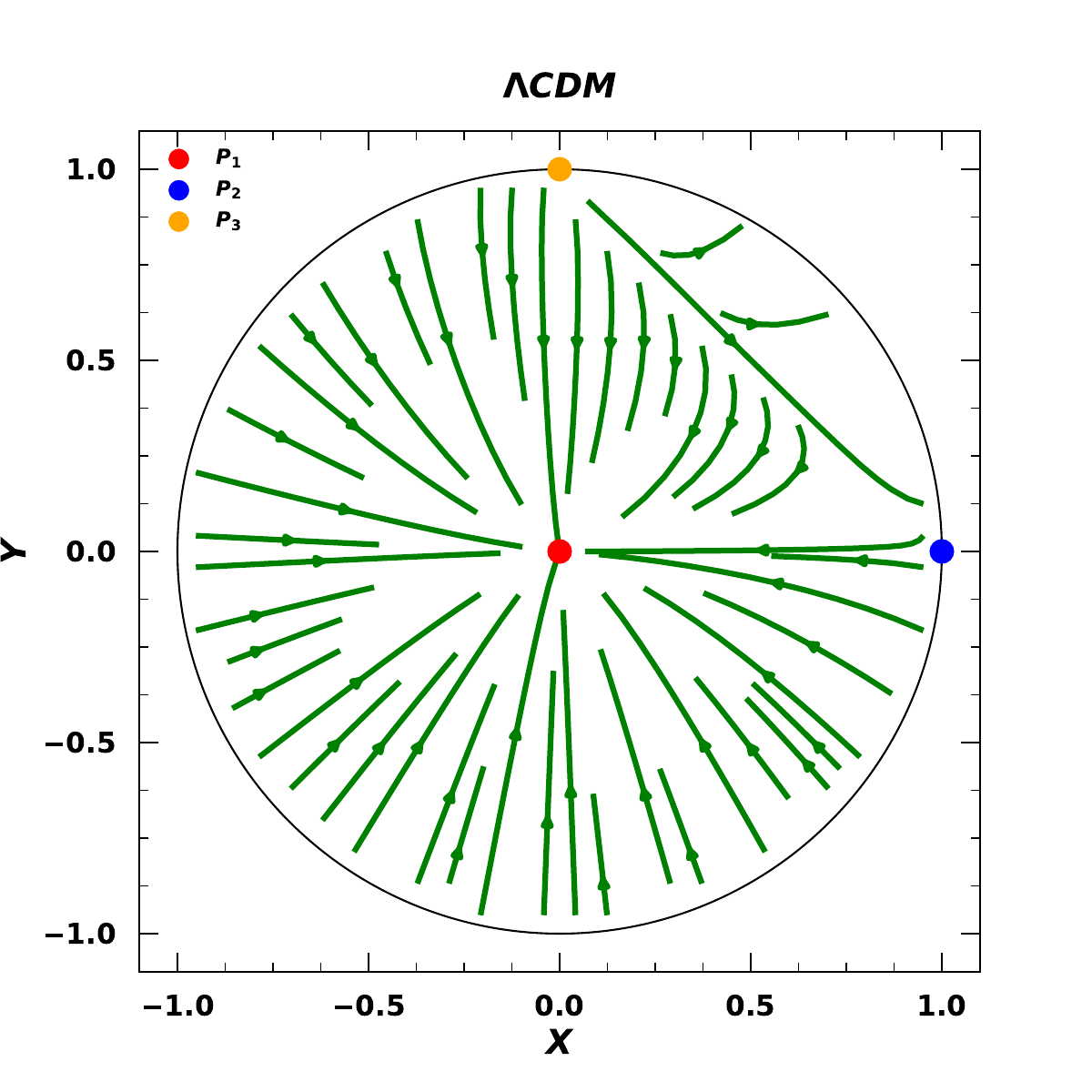}\hspace{0.25cm}
  \includegraphics[scale = 0.42]{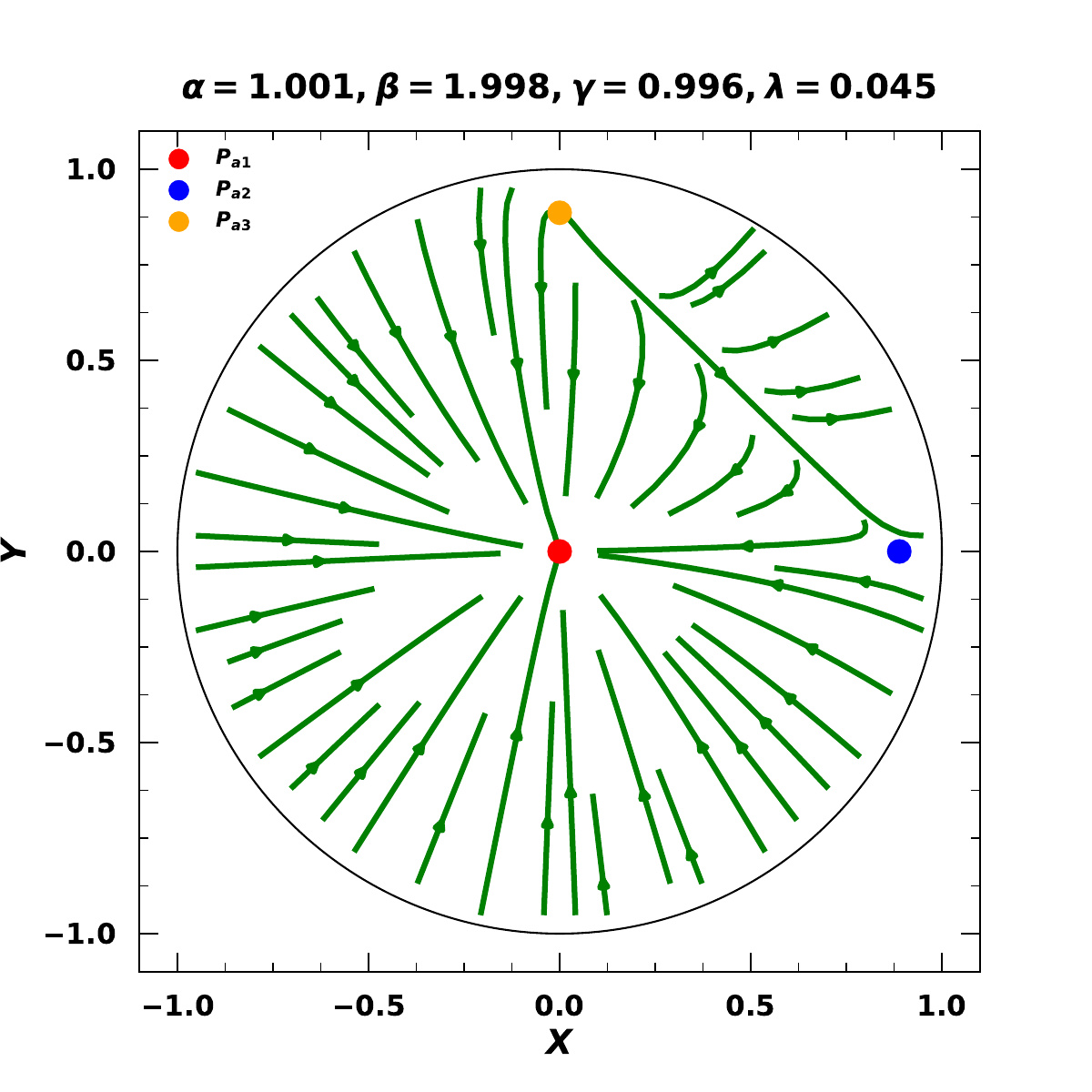}}
 \vspace{-0.25cm}
\caption{Phase space portraits for the $\Lambda$CDM model (left) and 
$f(R,T) = \alpha R + \beta f(T) $ model (right) for constrained values of the 
model parameters.}
\label{fig1}
\vspace{1cm}
\end{figure}

The $\omega_{eff}$ and $q$ can be  calculated from equations \eqref{Om_eff} 
and \eqref{dec} respectively. With 
the help of the dynamical system equations (\eqref{xdynamic} and 
\eqref{ydynamic}) we accomplish the fixed point analysis and results are 
compiled in Table \ref{table1} and the phase space portraits of $X$ versus 
$Y$ are shown for both the $f(R,T) = \alpha R + \beta f(T)$ and the standard
$\Lambda$CDM models in Fig.~\ref{fig1}. Here, in this work, we have used the
constrained values of the model parameters from the Ref.~\cite{Sarmah_24c}, 
where the authors comprehensively estimate these parameters by employing a 
powerful statistical technique known as Bayesian inference by using 
various cosmological observational data, including the data of Hubble 
parameter, BAO data, Pantheon plus data of Supernovae type IA, and CMB data 
etc and tabulate them for various $f(R,T)$ models.
\begin{center}
\begin{table}[!h]
\caption{The fixed point solutions for $f(R,T)= \alpha R + \beta f(T) $ model.}
\vspace{5pt}
\scalebox{0.95}{
\begin{tabular}{|c|c|c|c|c|c|c|}
\hline
\rule[1ex]{0pt}{2.5ex} Model & Fixed point &  $(X=\Omega_m,Y=\Omega_r)$ &  $Z=\Omega_\Lambda$ &  Eigenvalues  & $\omega_{eff}$   &  $q$ \\ 
\hline
\rule[1ex]{0pt}{2.5ex}& $P_1 $ & $\Big(0, 0\Big)$ & $1$ & $(-4,-3)$ & $-1.0$ & $-1.0$\\ 
\rule[1ex]{0pt}{2.5ex} {$\Lambda CDM$} &$P_2$ &$(1,0)$ & $0$ & $(3.0,-1.0)$ & $0$ & $0.5$\\
\rule[1.25ex]{0pt}{2.5ex} & $P_3$ &$\left(0,1\right)$ & $0$ & $(4.0,1.0)$ & $0.33$ &$1.0$\\ 
\hline
\rule[1ex]{0pt}{2.5ex}& $P_a1 $ & $\Big(0, 0\Big)$ & $0.7337$ & $(-3.8544,-3.1379)$ & $-0.99400$ & $-0.9909$\\ 
\rule[1ex]{0pt}{2.5ex} {$f(R,T) = \alpha R + \beta f(T)$}&$P_a2$ &$(0.8886,0)$ & $0.0079$ & $(3.1379,-0.7165)$&$-0.0271$ &$0.4613$\\
\rule[1.25ex]{0pt}{2.5ex} & $P_a3$ &$\left(0,0.8861\right)$ & $0.0429$ & $(3.8545,0.7166)$ & $0.2915$ &$0.9399$\\ 
\hline
\end{tabular}
}
\label{table1}
\end{table} 
\end{center}

The phase space portraits in Fig.~\ref{fig1} show the critical points 
indicating different phases of the BIII model Universe in the 
$f(R,T) = \alpha R + \beta f(T)$ model along with the $\Lambda$CDM
cosmological results. The phase portrait of our model is similar to the 
standard isotropic cosmology, which predicts the radiation-dominated phase, 
matter-dominated phase, and dark energy-dominated phase are hetero 
clinically connected. However, there are some subtle differences that have been 
observed in both the plots, too. The critical points for the BIII Universe are 
shifted to a less than unity state in contrast to the $\Lambda$CDM plot. These 
shifts occur due to the imposition of directional anisotropy in the BIII 
Universe. However, the track of the evolution of the Universe is intact as 
predicted by the standard cosmology, which means that there are 
radiation-dominated, matter-dominated, and dark energy-dominated phases, but 
in BIII Universe, this evolution track is influenced by an anisotropic effect. 
Further, from Table \ref{table1} we have observed that the $\omega_{eff}$ for 
the matter-dominated phase (i.e.~point $P_{a2}$) is non-zero but a very small 
negative value which may indicate a minute presence of dark energy influence 
in that phase. From the table, we also observed that the value $X+Y+Z \neq 1$ 
as the Friedmann equation is modified as equation \eqref{Fried1}. It 
indicates that for an anisotropic Universe, the total energy content consists 
of not only matter, radiation and dark energy, but there exist some other 
contributions too. The simplest candidate of it is basically the energy 
stored in the form of shear due to the presence of space-time deformation, as 
the Universe considered in anisotropic models has different expansion rates 
in different directions. Such findings are also recorded for the BI metric in 
$f(Q)$ gravity theory for different models that have been found in 
Ref.~\cite{Sarmah_2024}. More precisely, the term `shear energy' is 
associated with shear scalar ($\sigma^2$), which can be obtained from that 
shear tensor i.e.~${\sigma^2 = \frac{1}{2}\sigma^{\mu\nu}\sigma_{\mu\nu}}$ \cite{Sahoo_2016}. Further, the mathematical form of the shear tensor can be written as 
${\sigma_{\mu\nu} = \frac{1}{2}(u_{\alpha;\beta}+u_{\beta;\alpha})h^{\alpha}_{\mu}h^{\beta}_{\nu} - \frac{1}{3}u^{\alpha}_{;\alpha} h_{\mu\nu}}$. 
Here, $h_{\mu\nu}$ is the projection tensor and $u_{\alpha}$ is the four 
velocity in the comoving coordinates \cite{akarsu_2019}. For the BIII metric, 
the shear scalar further simplifies to $\sigma^2 = \frac{1}{2} (\sum_{i=1}^{3}H_{i}^2 - \frac{1}{3}\theta^2)$. Here, $\theta = (H_1+H_2+H_3)$ is the expansion 
scalar \cite{Sahoo_2016, Santhi_2018}. Thus, the shear scalar is directly connected with 
the anisotropic expansion rate of the Universe. Therefore, in this 
work, we have also interpreted the ${(1-X-Y-Z)}$ value of density as the shear 
energy stored in the form of spacetime deformation due to the presence of 
directional anisotropy with some model effect. 
\begin{figure}[!h]
\vspace{1cm}
\centerline{
  \includegraphics[scale = 0.42]{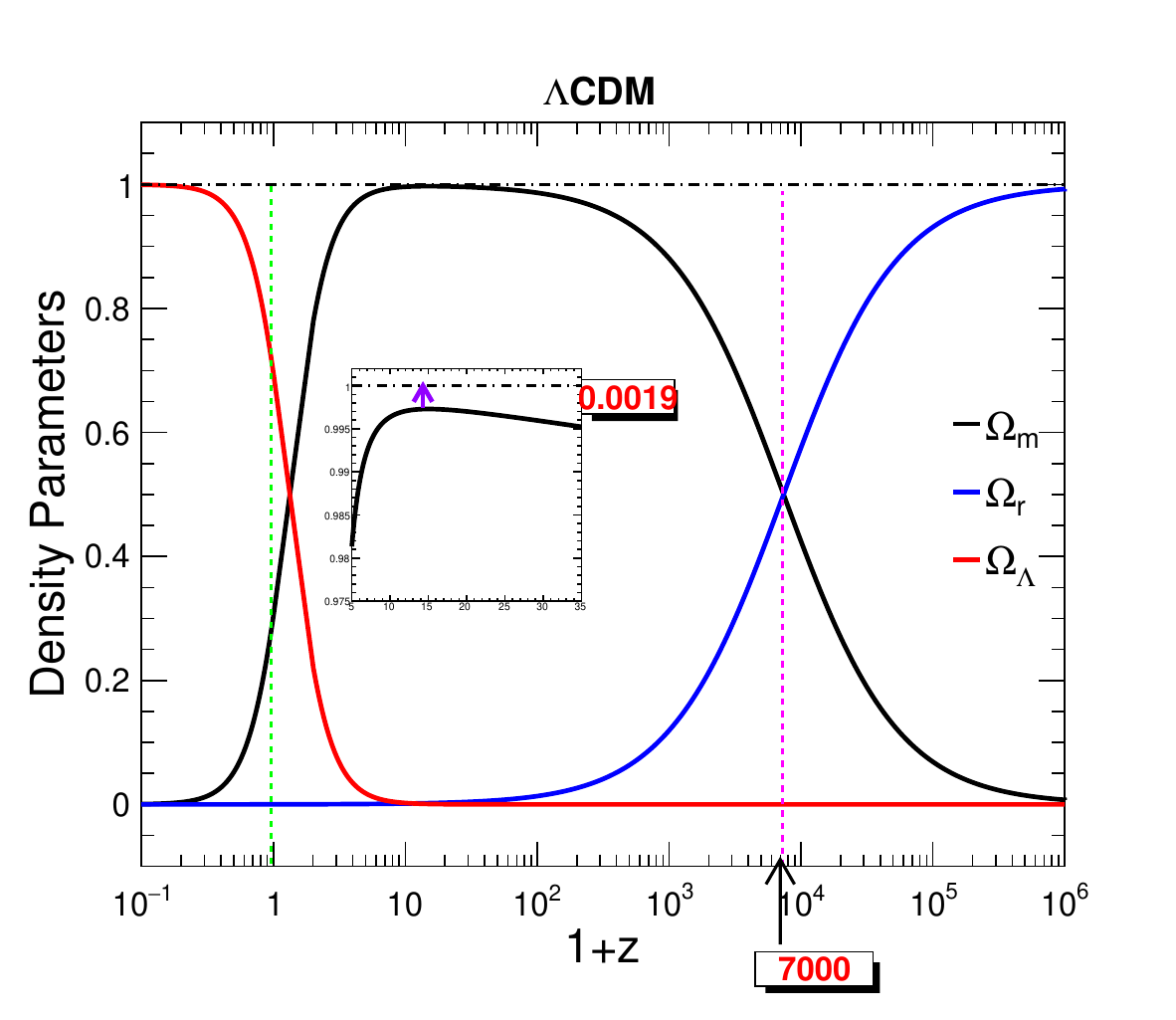}\hspace{0.25cm}
  \includegraphics[scale = 0.42]{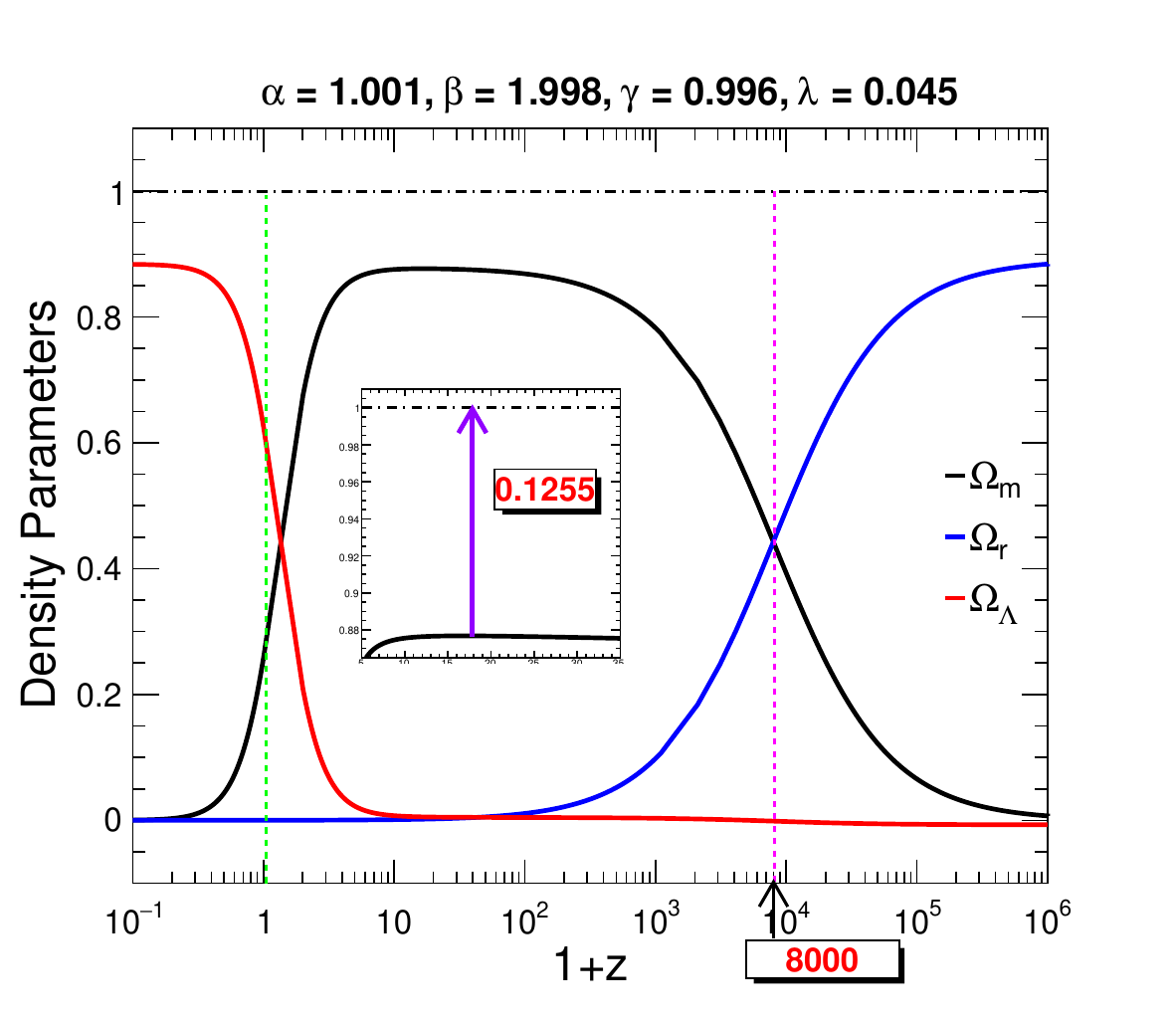}}
 \vspace{-0.25cm}
\caption{Variations of density parameters against the cosmological redshift 
for the $\Lambda$CDM model (left) and $f(R,T) = \alpha R + \beta f(T) $ model 
(right) for constrained values of the model parameters.}
\label{fig2}
\end{figure}

The evolutionary development of density parameters versus the cosmological 
redshift ($z$) for the considered $f(R,T)$ model with the constrained values 
of the model parameters has also been studied and compared the graphical 
results with the standard $\Lambda$CDM model findings in Fig.~\ref{fig2}. 
From the plots, one can find that there exists no ideal 
radiation, matter, or dark energy phases as we found them in the 
standard $\Lambda$CDM results, but show the dominance of radiation or matter 
or dark energy in those phases in BIII Universe as the density parameters 
never reached unity. The gap between the unity and the maximum value of any 
density parameter is mainly due to the presence of directional anisotropy 
along with the considered specific model. The anisotropy causes deformation 
in the spacetime, leading to shear energy stored in the spacetime fabric which 
also contributes in the Universe's total density. Apart from that, there is a 
shift in the value of $z$ from the standard $\Lambda$CDM results in BIII 
Universe for the considered model, when the matter-dominated phase had 
started dominating over the radiation-dominated phase. This shift is again 
due to the contribution of anisotropy and the effect of the model. Thus, we 
can remark that anisotropy may influence the evolutionary track of the 
BIII Universe.

\begin{figure}[!h]
\vspace{-0.5cm}
\centerline{
  \includegraphics[scale = 0.45]{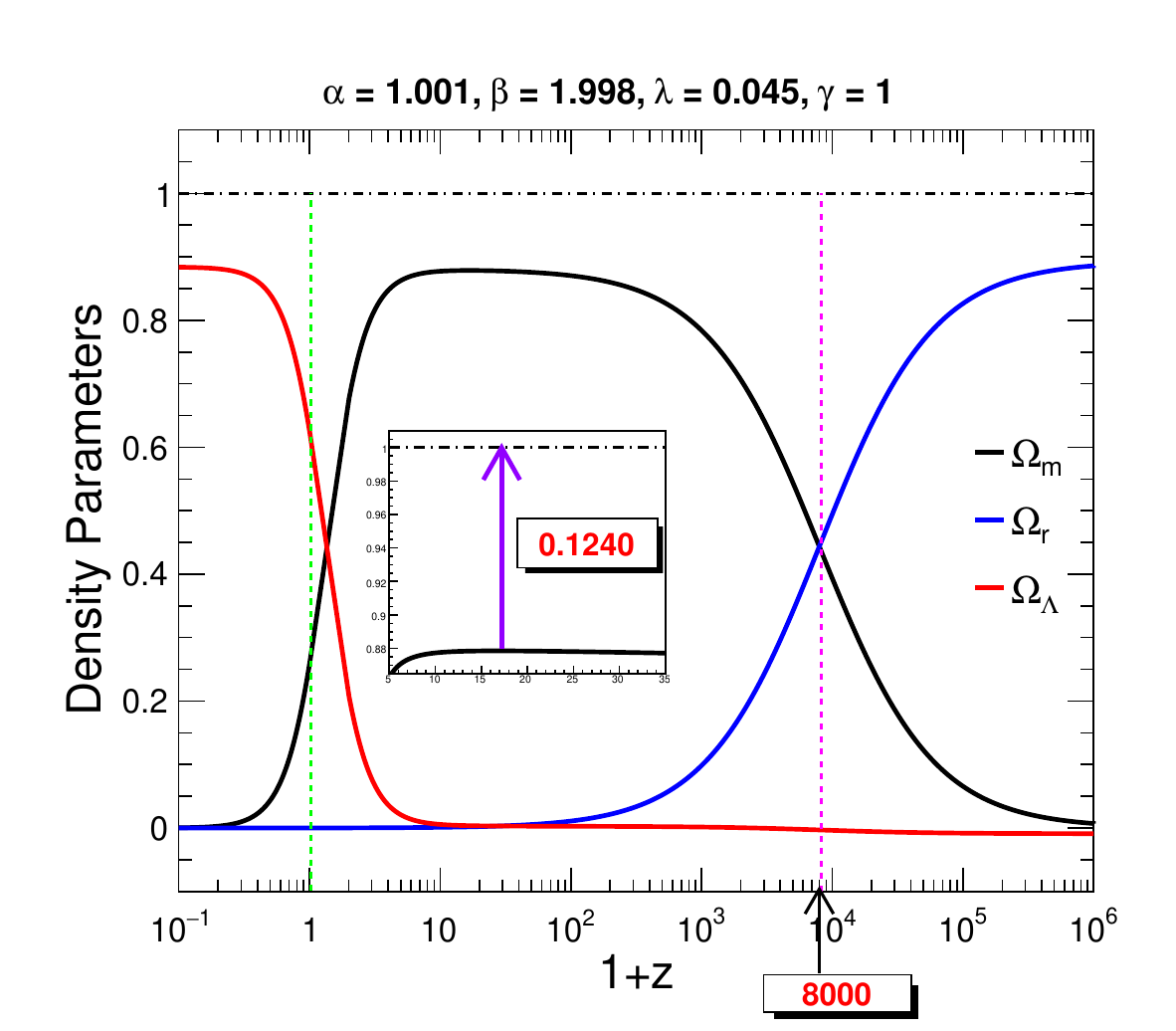}\hspace{0.25cm}
  \includegraphics[scale = 0.50]{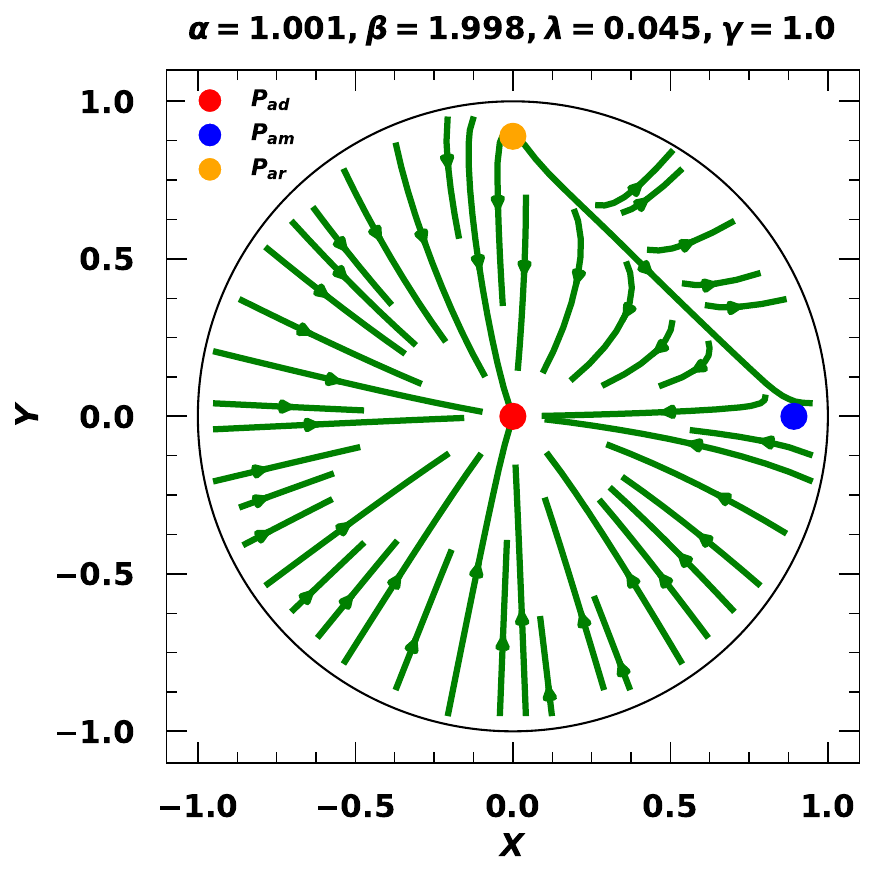}}
 \vspace{-0.25cm}
\caption{Variations of density parameters versus cosmological redshift for the 
$\Lambda$CDM model (left) and Phase space portrait for $f(R,T) = \alpha R + \beta f(T) $ model (right) for the constrained values of the model parameters with isotropic case.}
\label{fig2a}
\end{figure}
However, to quantify the effect of the model in the BIII Universe 
evolution, we have further tested the density parameter evolution and phase 
space portrait for the considered model by setting the $\gamma = 1$, which is 
nothing but the isotropic case as depicted in Fig.~\ref{fig2a}. 
From the figure, we have found that there is not much difference in the density 
parameter plots from Fig.~\ref{fig2}. There is a slight difference in the gap 
from unity to the maximum value of $\Omega_{m}$ between Fig.~\ref{fig2} and 
Fig.~\ref{fig2a}, which is around $0.0015$ and thus the effect is mainly due 
to the model. Hence, we can make a comment that anisotropy has a very minimal 
role in evolution, although it is not zero. Further, the point at which the 
transition from radiation-dominated phase to matter-dominated phase take place 
is the same for both isotropic and anisotropic cases, and hence no role of 
anisotropy on the shift of the transition point from the standard $\Lambda$CDM 
plot. This is purely a model effect. Apart from that, the phase space 
portrait of Fig.~\ref{fig1} (right) and Fig.~\ref{fig2a} are almost the same 
while following the heteroclinic path. 
However, the critical point in the isotropic case are $P_{ad} = (0, 0)$, 
$P_{am} = (0.8872, 0)$ and $P_{ar} = (0, 0.8838)$ which are slightly differ 
from the anisotropic case ($\gamma \neq 1$). In the anisotropic BIII metric, 
we have noticed that the fixed points are $P_{a1} = (0, 0)$, 
$P_{a2} = (0.8886, 0)$ and $P_{a3} = (0, 0.8861)$. With these observations, 
we can say that the model effect has a more dominant role in the evolution 
as compared to anisotropy.
  
\subsection{$\boldsymbol{f(R,T) = R + 2f(T)}$ Model}
\begin{figure}[!h]
\centerline{
  \includegraphics[scale = 0.42]{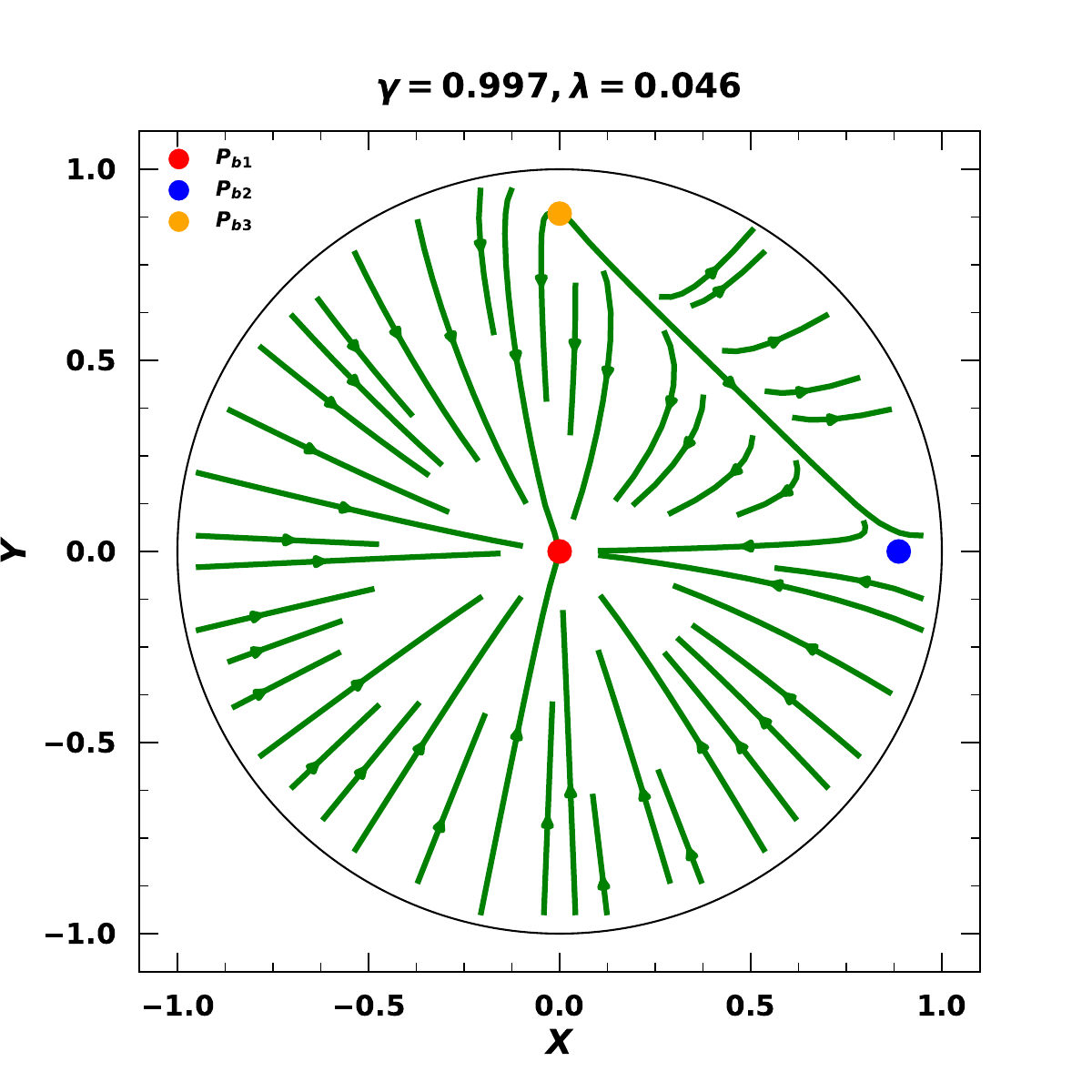}}
 \vspace{-0.25cm}
\caption{Phase space portrait for the $f(R,T) =  R + 2f(T)$ model for 
constrained values of the model parameters.}
\label{fig3}
\vspace{-0.5cm}
\end{figure}

The Friedmann like equation for the considered $f(R,T) = R + 2 f(T)$ model 
with $f(T) = \lambda T$ can be written from the equation \eqref{hub} as
\begin{equation}
\frac{3}{1+2\gamma}\left[(1+3\lambda)X + (1+\frac{4}{3}\lambda)Y + (1+8\lambda)Z + m^2S\right] = 1.
\end{equation}
Here, the variables $X,Y,Z$ and $S$ are as same as the previous 
model. Similarly, we can rewrite equation \eqref{e2} in terms of these 
dynamical variables with the form:
\begin{equation}
\frac{\dot{H}}{H^2} = -\,\frac{(2+\gamma)}{2}\left[(1+4\lambda)X + \frac{4}{3}(1+2\lambda)Y - \left\{ \frac{(1+2\gamma)}{3} - \frac{9}{(2+\gamma)^2} \right\}\right].
\end{equation}
Further, we have obtained the equations for the dynamical system i.e.~t $X'$ 
and $Y'$, which have the form:
  \begin{eqnarray}\label{dy1}
  X' = (2+\gamma)X \left[(1+2\lambda)X + \frac{4}{3}(1+2\lambda)Y -\left\{\frac{1+2\gamma}{3} - \frac{9}{(2+\gamma)^2} + \frac{3(1-2\lambda)}{(2+\gamma)(1-3\lambda)}\right\}  \right],\\[8pt]
\label{dy2}
  Y' = (2+\gamma)Y \left[(1+4\lambda)X + \frac{4}{3}(1+2\lambda)Y -\left\{\frac{1+2\gamma}{3} - \frac{9}{(2+\gamma)^2} + \frac{12(1-2\lambda)}{(2+\gamma)(3-4\lambda)}\right\}  \right].
  \end{eqnarray}
The $\omega_{eff}$ and $q$ can be derived 
from equations \eqref{Om_eff} and \eqref{dec} respectively, as we did for the 
previous model. From equations \eqref{dy1} and \eqref{dy2}, we analyze the
fixed points for this considered model and the results are compiled in 
Table \ref{table2}. Further, we plot the phase portrait of $X$ versus $Y$ for 
the $f(R,T) = R + 2\lambda T$ model as shown in Fig.~\ref{fig3}. 
Similar to the previous case, here also we have considered the constrained 
values of various parameters from Ref.~\cite{Sarmah_24c}.
\begin{center}
\begin{table}[!h]
\caption{The fixed point solutions for the $f(R,T)=  R + 2 f(T) $ model.}
\vspace{5pt}
\scalebox{0.95}{
\begin{tabular}{|c|c|c|c|c|c|c|}
\hline
\rule[1ex]{0pt}{2.5ex} Model & Fixed point &  $(X=\Omega_m,Y=\Omega_r)$ &  $Z=\Omega_\Lambda$ &  Eigenvalues  & $\omega_{eff}$   &  $q$ \\ 
\hline
\rule[1ex]{0pt}{2.5ex}& $P_{b1}$ & $\Big(0, 0\Big)$ & $0.7295$ & $(-3.8573, -3.1481)$ & $-0.9960$ & $-0.9940$\\ 
\rule[1ex]{0pt}{2.5ex} {$f(R,T) =  R + 2f(T)$}&$P_{b2}$ &$(0.8872,0)$ & $0.0085$ & $(3.1481, -0.7092)$&$-0.0291$ &$0.4577$\\
\rule[1.25ex]{0pt}{2.5ex} & $P_{b3}$ &$\left(0,0.8838\right)$ & $0.0437$ & $(3.8573, 0.7092)$ & $0.2885$ &$0.9346$\\ 
\hline
\end{tabular}
}
\label{table2}
\end{table}
\vspace{-0.8cm}
\end{center}

\begin{figure}[!h]
\centerline{
  \includegraphics[scale = 0.42]{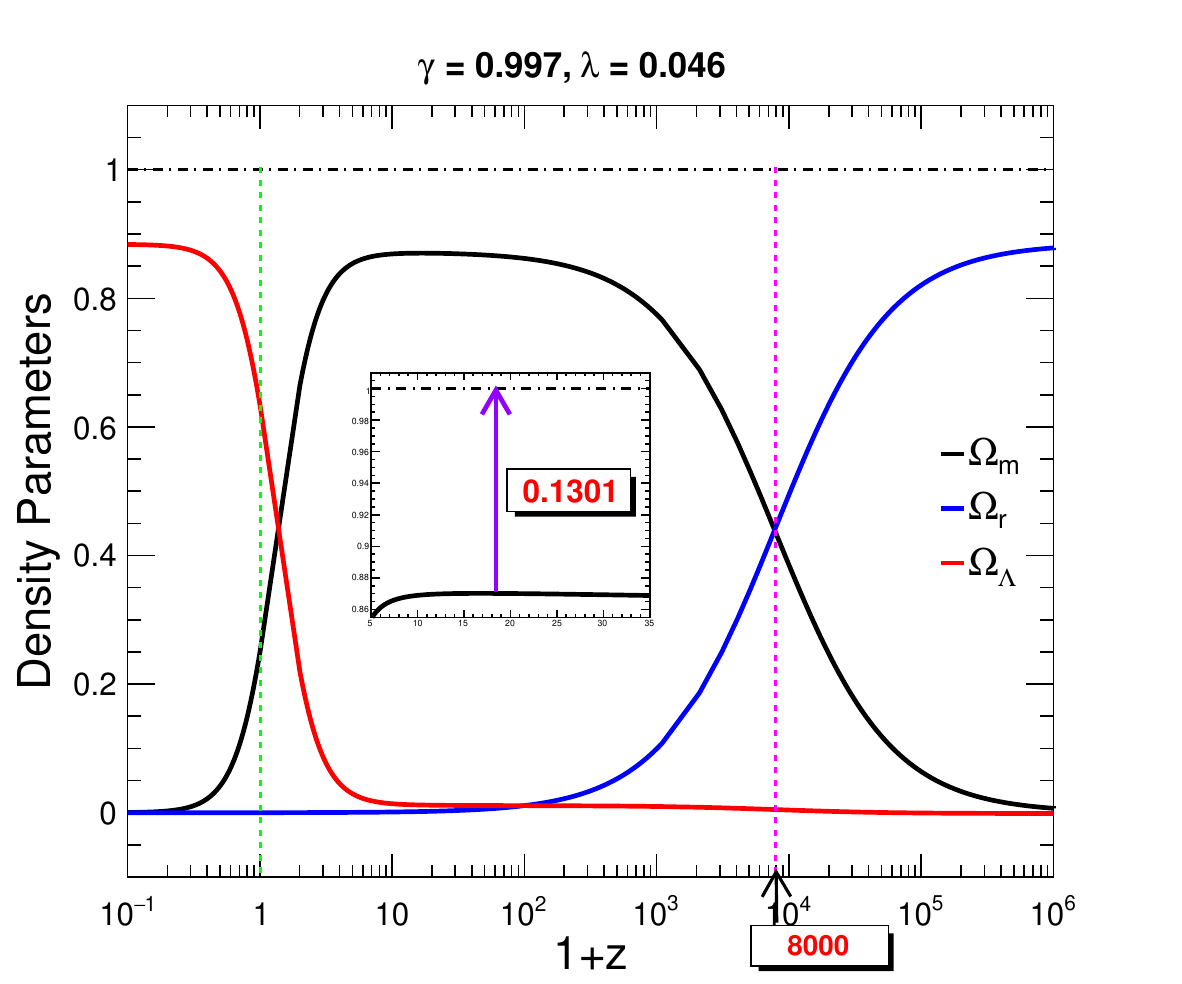}}
 \vspace{-0.25cm}
\caption{Density parameters versus cosmological redshift for the 
$f(R,T) =  R + 2f(T)$ model for constrained values of the model parameters.}
\label{fig4}
\end{figure}

From Fig.~\ref{fig3} and Table \ref{table2}, like in the previous model we 
can observe the shift of fixed points from the  $\Lambda$CDM model 
for this model too. As we have previously claimed, the anisotropy has the role 
of shifting these points from the unit value here in this model too. Further, 
there is a very small negative value of $\omega_{eff}$ observed in the 
matter-dominated phase, which indicates a very small role of dark energy in the 
matter-dominated region of the evolution in the BIII type of Universe. Apart 
from that, we observe a similar hetero clinic sequence of evolution of the 
BIII Universe like standard cosmology, i.e.~Universe evolved through
radiation-dominated phase to matter-dominated phase and then to dark 
energy-dominated phase ($P_{b3} \rightarrow P_{b2} \rightarrow P_{b1}$). 
Moreover, like in the 
previous model analysis, we see that $\boldsymbol{1-X-Y-Z \neq 0}$, and this non-zero 
contribution is due to anisotropy and the effect of the model
, which we have already discussed in our previous model. The presence of these 
effects can be observed in all three phases. Thus, a prominent role of model 
and a minimal role of anisotropy are present in the entire development of the 
various phases of the 
Universe, as claimed in the previous model analysis. As in the case of the 
previous model, we also study the evolution of the density parameters with 
cosmological redshift in this model, which is shown in Fig.~\ref{fig4}. 
The model shows almost similar results as in the previous case and clearly 
indicates the influence of the model and the anisotropy in the overall 
Universe's evolution.
\subsection{$\boldsymbol{f(R,T) =f_1 (R)+ f_2(R)f_3(T)}$: A Special Case}\label{5}
\begin{figure}[!h]
\centerline{
\includegraphics[scale = 0.42]{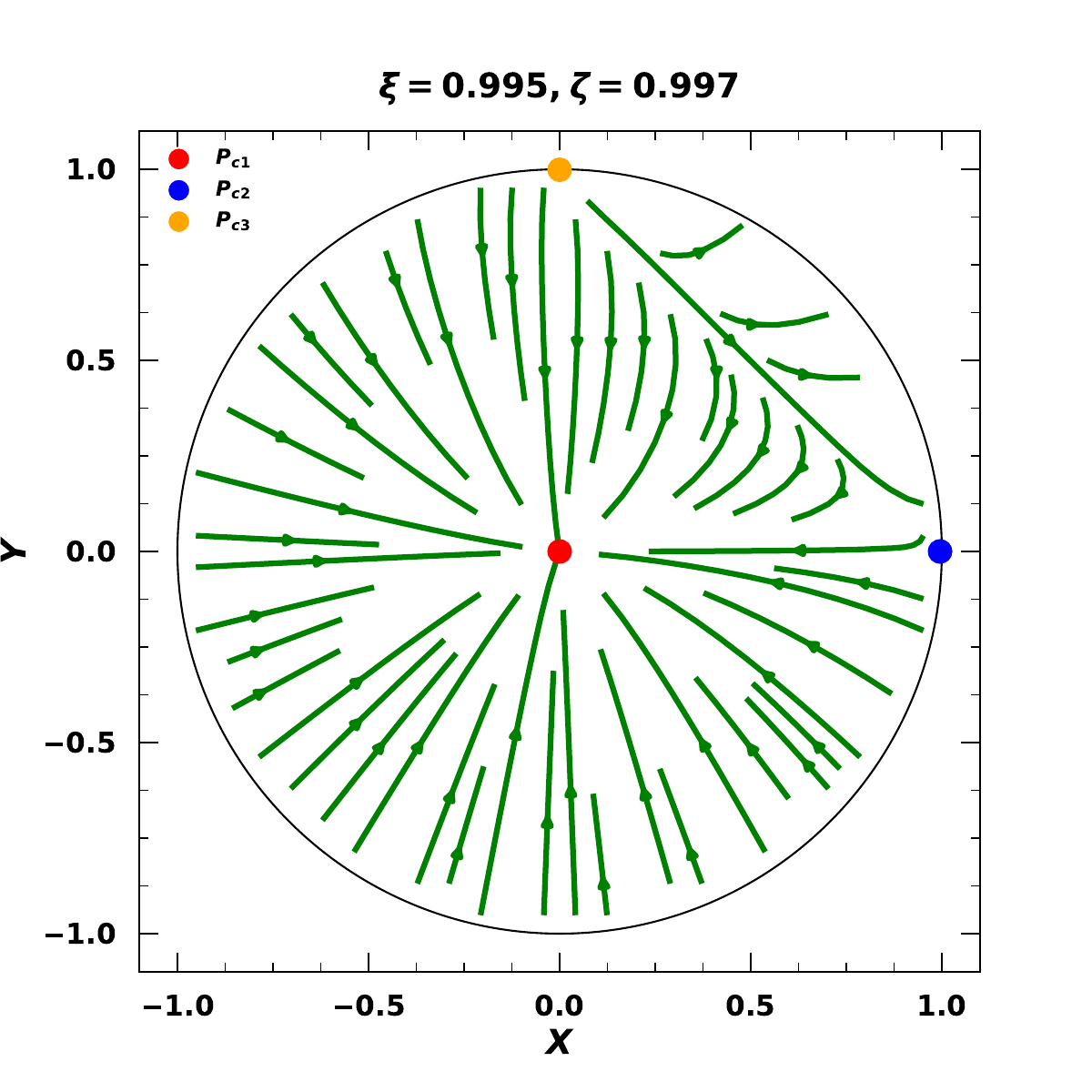}}
\vspace{-0.25cm}
\caption{Phase space portrait for the $f(R,T) = (\zeta + \eta\, \tau\, T)R$ 
model with the constrained values of the model parameters.}
\label{fig5}
\end{figure}

Here, we are interested in studying a special model 
$f(R,T) =f_1 (R)+ f_2(R)f_3(T)=(\zeta + \eta\, \tau\, T)R $ {in which} 
${f_{1}(R)= \zeta R,}$ ${f_{2}(R) = \tau R, f_{3}(T) = \eta T}$. 
Our interest in this model is based on the fact that an inconsistency 
(sharp discontinuity in the radiation-dominated region) from the standard 
cosmological evolution as predicted by the $\Lambda$CDM model has been 
recorded for this non-linear model in Ref. \cite{Sarmah_24c}. The modified 
Einstein field equations for the chosen model may have the form 
\cite{Sarmah_24c}:
\begin{eqnarray}\label{Fei}
3H^2 = \frac{(2+\xi)^2}{3(1+2\xi)}\left[\frac{\rho+ \eta \tau R(\rho-p)}{\zeta +\eta \tau T} +\frac{m^2}{a^{\frac{6}{2+\xi}}}\right],\\[8pt]
3H^2+ \frac{2}{3}(2+\xi) \dot{H}  = -\frac{(2+\xi)^2}{9}\left[\frac{(1+2\eta \tau R)p}{\left(\zeta+\eta \tau T\right)}-\frac{m^2}{a^{\frac{6}{2+\xi}}}\right].\label{Feii}
\end{eqnarray}
Here $\zeta, \eta, \tau$ are model parameters, and $\xi$ is the directional 
anisotropy parameter ($\gamma$ as in previous models). As previously mentioned 
the conventional energy-momentum tensor is not conserved in $f(R,T)$ 
theory of gravity, thus the modified form of conservation condition, 
$\nabla_{\mu}T^{\mu\nu}_{eff} = 
\nabla_{\mu}T^{\mu\nu}+ \nabla_{\mu}{\tilde T^{\mu\nu}} = 0$ 
has been applied to obtain the required form of the continuity equation. Here 
in this model,
$T^{00}_{eff} = \frac{\rho+\eta \tau R (\rho - p)}{ \left(\zeta+\eta \tau T\right)}$, and $T^{ii}_{eff} = -\,g^{ii}\frac{(1 + 2\,\eta \tau R)p}{ \left(\zeta+\eta \tau T\right)}$, where $i=(1,2,3)$. The continuity 
equation obtained here now can be written as \cite{Sarmah_24c}
\begin{align}\label{cont_3}
\frac{3 \dot{a}}{a}& \left[{(\omega +1)\rho-\frac{2 \eta  \tau \rho^2 (1-3 \omega)^2}{\zeta +2 \eta  \tau  (5 \omega -1)\rho}}\right]+ \frac{2 \eta ^2 \tau ^2 (1-\omega) (1 -3 \omega) \left(5 \omega-1\right)\dot{\rho}\rho^2}{\{\zeta +2 \eta  \tau  (5 \omega -1) \rho\}^2}\nonumber\\[5pt]
&-\frac{2\eta  \tau  (1-3 \omega )\left(1-\omega\right)\rho \dot{\rho}}{\zeta +2 \eta  \tau  (5 \omega -1)\rho}+\dot{\rho} - \frac{\eta  \tau  \left(3 \omega -1\right)\dot{\rho} \left(\rho-\frac{\eta  \tau  (1 -3 \omega) (1-\omega)\rho^2}{\zeta +2 \eta  \tau  (5 \omega -1)\rho}\right)}{\{\zeta +\eta  \tau  (3 \omega -1)\rho\}}=0.
\end{align}
The solutions of this complex form of the continuity equation can be obtained by 
splitting the equation into three components, which correspond to three 
density  $\rho_m$, $\rho_r$ and $\rho_{\Lambda}$ for the matter state
($\omega = 0$), radiation state ($\omega = \frac{1}{3}$) and dark energy state
($\omega = -1$)  respectively of the Universe, and then solving each 
equations give three separate equations as given below \cite{Sarmah_24c}:
\begin{align}\label{solm}
\rho_m & = \frac{\zeta  a^3+3 \zeta  \eta  \tau  e^{c_1 \zeta }\pm \zeta  \sqrt{-6 \eta  \tau  a^3 e^{c_1 \zeta }+a^6+\eta ^2 \tau ^2 e^{2 c_1 \zeta }}}{2 \left(3 \eta  \tau  a^3+2 \eta ^2 \tau ^2 e^{c_1 \zeta }\right)},\\[8pt]
\label{solr}
\rho_r& = \frac{c_1}{a(t)^4} = \rho_{r0} a^{-4},\\[8pt]
\label{soll}
\rho_\Lambda & =  \log(\text{constant~term}) = \rho_{\Lambda0}.
\end{align}
By neglecting the higher order terms of $\eta$ and $\tau$ as we are expecting 
small deviations of the model's result from the GR result, the reduced form 
of equation \eqref{solm} can be written for the negative sign before the 
square root term as
\begin{equation}\label{solmn}
\rho_m = \zeta \rho_{m0}\, a^{-3}.
\end{equation}
These equations are now ready for obtaining dynamical system equations. Now 
from equation \eqref{Fei}, the Friedmann-like equation can be obtained as
\begin{equation}\label{1e}
\frac{(2+\xi)^2}{3(1+2\xi)}\left[\frac{1}{(\zeta + \eta \tau T)}\left\{(1+ \eta \tau R)X +  (1+\frac{2}{3}\,\eta \tau R)Y + (1+2\eta \tau R)Z\right\} + m^2 S \right] = 1.
\end{equation}
Again, from equation \eqref{Feii} we can obtain 
\begin{equation}\label{2e}
\frac{\dot{H}}{H^2} = -\frac{(2+\xi)}{2}\left[\frac{1}{(\zeta + \eta \tau T)}\left\{\frac{1}{3}(1+2\eta \tau R)Y - (1+2\eta \tau R)Z \right\}-m^2 S\right] - \frac{9}{2(2+\xi)}.
\end{equation}
With the help of equation \eqref{1e} and \eqref{2e} we construct the dynamical 
system equations as
\begin{eqnarray}\label{3e}
X' = X\left[\frac{(2+\xi)}{(\zeta + \eta \tau T)}\left\{(1+\,\eta \tau R)X + \frac{4}{3}\,(1+\eta \tau R)Y\right\} -3 + \frac{6(1-\xi)}{(2+\xi)}\right],\\[8pt]
%
\label{4e}
Y' = Y\left[\frac{(2+\xi)}{(\zeta + \eta \tau T)}\left\{(1+\,\eta \tau R)X + \frac{4}{3}\,(1+\eta \tau R)Y\right\} -4 + \frac{6(1-\xi)}{(2+\xi)}\right].
\end{eqnarray}
Both equations \eqref{3e} and \eqref{4e} are not only dependent on dynamical 
variables $X$ and $Y$, like previously mentioned models but also depend on 
other variables $R$ and  $T$ which them self are the functions of $X,Y,Z$ and $S$. Thus, the situations are more complicated in this model. However, from 
Ref.~\cite{Sarmah_24c}, we found that the constrained values of 
$\eta = 0.000015^{+0.0000017}_{-0.0000015}$ and 
$\tau =0.00000036^{+00000012}_{-0.0000016}$ are very small and thus, we 
have considered $(1+ \eta \tau R) \sim 1$ and 
$(\zeta + \eta\, \tau\, T) \sim \zeta$. With these approximations, we can 
rewrite the equation \eqref{3e} and \eqref{4e} as 
\begin{eqnarray}\label{5e}
X' = X\left[\frac{(2+\xi)}{\zeta}\left\{X + \frac{4}{3}Y\right\} -3 + \frac{6(1-\xi)}{(2+\xi)}\right],\\[8pt]
\label{6e}
Y' = Y\left[\frac{(2+\xi)}{\zeta}\left\{X + \frac{4}{3}Y\right\} -4 + \frac{6(1-\xi)}{(2+\xi)}\right].
\end{eqnarray}
We have compiled the fixed-point solutions for the considered model in Table 
\ref{table3} and drawn the phase portrait for $X$ versus $Y$ for the 
constrained values of the model parameters from Ref.~\cite{Sarmah_24c} in 
Fig.~\ref{fig5}. $\omega_{eff}$ and $q$ in this table are obtained by using 
equations \eqref{Om_eff} and \eqref{dec} respectively with replacing 
$\gamma$ by $\xi$ as directional anisotropic parameter. 
\begin{center}
\begin{table}[!h]
\caption{The fixed point solutions for $f(R,T)= (\zeta + \eta\, \tau\ T)R $ 
model with the constrained values of model parameters.}
\vspace{5pt}
\scalebox{0.95}{
\begin{tabular}{|c|c|c|c|c|c|c|}
\hline
\rule[1ex]{0pt}{2.5ex} Model & Fixed point &  $(X=\Omega_m,X=\Omega_r)$ &  $ Z_{t}$ &  Eigenvalues  & $\omega_{eff}$   &  $q$ \\ 
\hline
\rule[1ex]{0pt}{2.5ex}& $P_{c1}$ & $\Big(0, 0\Big)$ & $0.9969$ & $(-3.9899, -2.9899)$ & $-0.9966$ & $-0.9949$\\ 
\rule[1ex]{0pt}{2.5ex} {$f(R,T) = (\zeta + \eta \tau T)R$}&$P_{c2}$ &$(0.9953,0)$ & $0.0016$ & $(2.98998, -1.0)$&$-0.0016$ &$0.5$\\
\rule[1.25ex]{0pt}{2.5ex} & $P_{c3}$ &$\left(0,0.9986\right)$ & $-0.0017$ & $(3.98998, 1.0)$ & $0.3344$ &$1.005$\\ 
\hline
\end{tabular}
}
\label{table3}
\end{table} 
\end{center}

From Table \ref{table3}, it is found that there are three critical points 
$P_{c1}, P_{c2}$ and $P_{c3}$ corresponding to three different phases of the 
Universe which are mainly dark energy, matter, and radiation era as predicted 
by standard 
cosmology, and the phase portrait in Fig.~\ref{fig5} shows the hetero clinic 
transition from $P_{c3}\rightarrow P_{c2} \rightarrow P_{c1}$ which is again 
predicted by the standard cosmology. However, the values of $X$ 
and $Y$ are found to be close to standard $\Lambda$CDM model values. In the tbale  \ref{table3} 
$Z_t$ refers to the sum of the contribution of $Z$ and the anisotropic effect 
due to $S$.  However, for the matter-dominated phase (i.e. $P_{c2}$) the 
values of $\omega_{eff}$ is non-zero and negative, which does not match 
standard cosmology, {as the negative value of the equation of state 
leads to negative pressure and hence supports accelerated expansion, which is 
a contrasting scenario from the standard cosmology in the matter dominated 
phase}. Thus, it further supports the claim of non-suitability of this model 
to study cosmological evolution as claimed by Ref.~\cite{Sarmah_24c}. Apart 
from that, at $P_{c3}$  $Z_t$ gives a negative value which has no physical 
meaning and thus raises the 
question of the physical viability of the model. Therefore, even though the 
model predicts the standard cosmological transition path as predicted by the
$\Lambda$CDM model, its results are not physically viable. Further, the
inconsistency in the density parameters and the other cosmological parameter 
values from standard cosmology in this special case is mainly due to this 
considered model itself, since in the study of the previous two models, such 
inconsistency has not been observed, and they provide results within the 
physically viable range. Thus, the model appears to be unsuitable for 
cosmological study in the BIII Universe. 

Since in the first two models, we have observed that the model has more 
dominant role than anisotropy and thus, the strange unphysical results observe 
in the special case  is may also be due to the model and hence further
strengthens our claim of non-suitability of the considered model. Again, the 
results are not physically viable, hence we have restricted ourselves to further 
studies.
    
\section{Conclusions}\label{6}
To understand the Universe as a dynamical system and check its stability by 
means of critical-point analysis, we have considered the BIII metric with 
three different models within the $f(R,T)$ theory of gravity. We have also 
tried to understand the role of anisotropies in the cosmic evolution of the 
BIII Universe. Starting with the general form of the field equations for the 
$f(R,T)$ theory of gravity and related equations and expressions in Section 
\ref{2}. We then further extended these equations for the BIII metric in 
{three} different $f(R,T)$ models in  Section \ref{3}.
 
Section \ref{4} covers the  dynamical system analysis for the 
$f(R,T) = (\alpha R + \beta T)$ and $f(R,T) = R + 2 f(T)$ models. Here, we 
have constructed the system of equations by using field equations for both 
models by considering mainly the density parameters as the dynamical 
variables. Apart from that, we have also derived the effective equation of 
state ($\omega_{eff}$) and deceleration parameter ($q$), and carried forward 
the stability analysis of the considered dynamical system for both the models. 
We have compiled our results in Table \ref{table1} for the 
$f(R,T) = \alpha R + \beta f(T)$ model and similarly the results for the 
$f(R,T) = R + 2 f(T)$ model in Table \ref{table2}. The respective phase 
portraits corresponding to  $f(R,T) = \alpha R + \beta f(t)$ and 
 $f(R,T) = R + 2 f(T)$ models are shown accordingly in Fig.~\ref{fig1} and 
Fig.~\ref{fig3}. We have further compared our findings with the $\Lambda$CDM 
model results in Table \ref{table1} and Fig.~\ref{fig1}. From our 
analysis, we have found that both models support the standard evolution 
process as predicted by $\Lambda$CDM cosmology, but with some model and 
anisotropic effects. The shifting of critical points from unity value in phase 
portraits in Fig.~\ref{fig1} (right) and Fig.~\ref{fig3} indicate that even 
though the radiation, matter and dark energy phases are connected 
through heteroclinic path in these two $f(R,T)$ models like $\Lambda$CDM 
cosmology, the 
highest value of the dynamic density parameters always lower than the unity 
due to the presence of anisotropic background in the BIII metric and for the 
considered specific model. The role of anisotropy in the evolution of the 
Universe can also be seen through Fig.~\ref{fig2} and Fig.~\ref{fig4}, 
where the evolution of density parameters versus cosmological redshift ($z$) 
has been shown and also compared it with $\Lambda$CDM results in 
Fig.~\ref{fig2} (left). These figures clearly show the non-unity of the 
maximum value of density parameters, hence supporting the dominated phases of 
radiation or matter, or dark energy rather than the pure phases due to the 
effect of the model and the anisotropies. However, from Fig ~\ref{fig2a} we 
have observed that the effect of the model is more prominent than the 
anisotropy.
   
In section \ref{5} we have discussed a special model $f(R,T) = 
(\zeta + \eta\, \tau\, T)R$ that we have studied in our previous work 
\cite{Sarmah_24c}. In that work, we found some inconsistency in this 
considered model from standard $\Lambda$CDM prediction. We found a sharp 
discontinuity of cosmological parameters in radiation-dominated era for $f(R,T) = (\zeta + \eta\, \tau\, T)R$ model, and hence we 
have commented on this model as not suitable for cosmological study in the 
BIII metric. To strengthen our claim, we also studied the model in the 
BIII metric in the current framework, considering it in the dynamical system, 
constructed a dynamical set of equations, similarly done in the previous models 
and compiled the results in Table \ref{table3}. We have also drawn phase 
portraits of $X = \Omega_{m}$ against $Y = \Omega_{r}$ in Fig.~\ref{fig5}. 
From Table \ref{table3} for the constrained set of models and cosmological 
parameters we have found that the value of $Z_t$ at ($P_{c3}$) i.e. radiation  
dominated phases is negative, which is physically impossible. Apart from that, 
we have  also observed non zero but negative value of $\omega_{eff}$ at
$P_{c2}$ i.e. matter matter-dominated region, which indicates the deviations 
from standard cosmological predictions and evolutions of the Universe. Thus 
we can make remark on this  model is that the model may have some problems in 
dealing with the radiation-dominated phase. Hence, as we claimed in our 
previous work \cite{Sarmah_24c}, we again claimed  that this model has not 
passed the test of dynamical system analysis and thus is not suitable for 
cosmological study for the BIII metric.
  
Lastly, we would like to point out that for this study in $f(R,T)$ gravity 
theory, we have considered all the models from our previous work
\cite{Sarmah_24c}. The first two models i.e.~$f(R,T) = \alpha R + \beta f(T)$ 
and $f(R,T) = R + 2 \lambda f(T)$ show consistency with the $\Lambda$CDM 
cosmology with some model and anisotropic effects. However, the third model 
shows inconsistency with $\Lambda$CDM cosmology in the matter-dominated 
era and hence we again claim that this model has issues in studying Bianchi 
cosmology as it is not a physically compatible model as mentioned already.  
With an excellent agreement with observational data and standard cosmological 
results, the first two considered models have been passed handsomely and they 
could explain the effect of anisotropies in the cosmic evolution and may 
provide valuable information on anisotropies in Universe's early stages.  
However, to confirm the existence of the anisotropy and their role in that 
early stage of the Universe, we need more observational evidence of the 
early stages of the Universe. This mystery could be unfolded by the Thirty Meter Telescope (TMT)\cite{tmt}, Extremely Large 
Telescope \cite{elt}, CTA \cite{cta}, and other upcoming projects. We hope that 
these upcoming projects will help to understand the development of 
the early stages of the Universe.

\appendix
\section{Derivation of different components of $\tilde{T}_{\mu\nu}$}
Here, we have shown the derivation of the components of $\tilde{T}_{\mu\nu}$ 
for $f(R,T) = \alpha R + \beta f(T)$ and $f(R,T) = R + 2 f(T)$ models. In this 
context we consider $f(T) = \lambda T$, thus the derivative 
$f_{T}(T) = \lambda$. Hence from equation \eqref{EMTt} we can write 
\begin{equation}
\kappa \tilde{ T}_{\mu \nu} = \lambda \beta \left [T_{\mu\nu} + \left \{p + \frac{1}{2} T\right\}g_{\mu\nu}\right].
\end{equation}
Since, $T = -\rho + 3p,$ above equation can be written as
\begin{equation}
\tilde{T}_{\mu\nu} = \frac{\lambda \beta}{\kappa} \left[ T_{\mu\nu} + \frac{1}{2} \left\{-\rho + 5p\right\}g_{\mu\nu} \right].
\end{equation}
Again, $g_{00} = -1, g_{11} = 
a_1^2, g_{22} = a_2^2 e^{-2mx}, g_{33} = a_3^2$, and 
$T_{00} = \rho, T_{11} = a_1^2 p, T_{22} = a_2^2 e^{-2mx}p, 
T_{33} = a_3^2 p,$ and also for $\kappa = 1$, 
we can write,
\begin{eqnarray}
\tilde{T}_{00} = \frac{\lambda \beta}{2}(3\rho-5p),\\
\tilde{T}_{11} = \frac{a_1^2\lambda \beta}{2}(-\rho+7p),\\
\tilde{T}_{22} = \frac{a_2^2 e^{-2mx}\lambda \beta}{2}(-\rho+7p),\\
\tilde{T}_{33} = \frac{a_3^2\lambda \beta}{2}(-\rho+7p).
\end{eqnarray} 
A similar approach would give the different components of  $\tilde{T}_{\mu\nu}$ for the model III, i.e. $f(R,T) = (\zeta+\eta \tau T)R$ also.

\section{Derivation of equations (39) and (40)}
As previously defined in Section \ref{4}, $X = \frac{\rho_m}{3H^2}$, $Y = \frac{\rho_r}{3H^2}$, $Z = \frac{\rho_{\Lambda}}{3H^2}$ and $S = 1/3H^2 a^{\frac{6}{2+\gamma}}$ 
are dynamical variables.
Further, $X' = \frac{dX}{dN}$ with $N = \log a$. 
Thus,
\begin{align}\label{a1}
X'& = \frac{dX}{dN} = \frac{1}{H}\frac{dX}{dt}= \frac{\dot{\rho_m}}{3H^3} - 2{\frac{\rho_m}{3H^2}} \frac{\dot{H}}{H^2},\\[8pt]
\label{a2}
Y' & = \frac{dY}{dN} = \frac{1}{H}\frac{dY}{dt} = \frac{\dot{\rho_r}}{3H^3} - 2{\frac{\rho_r}{3H^2}} \frac{\dot{H}}{H^2}.
\end{align}
Now, from equation \eqref{cont}, we can write $\dot{\rho_m} = \frac{-3H\rho_m (1-\lambda\beta)}{1-\frac{3\lambda\beta}{2}}$ 
and $\dot{\rho_r} = \frac{-4H\rho_r (1-\lambda\beta)}{1-\frac{2\lambda\beta}{3}}$ for 
$\omega = 0$ and $\omega = \frac{1}{3}$ respectively, and also using the equation \eqref{e2},  
equations \eqref{a1}  and \eqref{a2} can be transformed as
\begin{eqnarray}
X' = \frac{2+\gamma}{\alpha}X \left[(1+\lambda\beta)X + \frac{4}{3}(1+\lambda \beta)Y - 3\alpha\left\{\frac{1+2\gamma}{(2+\alpha)^2} - \frac{3}{(2+\gamma)^2} + \frac{2(1-\lambda \beta)}{(2+\gamma)(2-3\lambda\beta)}\right\} \right],\\[8pt]
Y' = \frac{2+\gamma}{\alpha}Y \left[(1+\lambda\beta)X + \frac{4}{3}(1+\lambda \beta)Y - 3\alpha\left\{\frac{1+2\gamma}{(2+\alpha)^2} - \frac{3}{(2+\gamma)^2} + \frac{4(1-\lambda \beta)}{(2+\gamma)(3-2\lambda\beta)}\right\} \right].
\end{eqnarray}
In similar way we can obtain equations (\ref{dy1}, 
\ref{dy2}) and (\ref{3e}, \ref{4e}). 
 
\section*{ACKNOWLEDGEMENT}
UDG is thankful for the Visiting Associateship of the  Inter-University Centre for Astronomy and Astrophysics (IUCAA), Pune, India .

\bibliographystyle{apsrev}

\end{document}